\documentclass[12pt,preprint]{aastex}
\usepackage{epsf}

\begin{document} 

\title{Terrestrial Planet Formation 
I. The Transition from Oligarchic Growth to Chaotic 
Growth\footnote{Calculations reported here used the
`hydra' cluster run by the Computation Facility of the
Harvard-Smithsonian Center for Astrophysics}}

\author{Scott J. Kenyon}
\affil{Smithsonian Astrophysical Observatory, 60 Garden Street, 
Cambridge, MA 02138, USA; e-mail: skenyon@cfa.harvard.edu}
\author{and}
\author{Benjamin C. Bromley}
\affil{Department of Physics, University of Utah, 201 JFB, Salt Lake City, 
UT 84112, USA; e-mail: bromley@physics.utah.edu}

\begin{abstract}
We use a hybrid, multiannulus, $n$-body--coagulation code 
to investigate the growth of km-sized planetesimals at
0.4--2 AU around a solar-type star.  After a short runaway 
growth phase, protoplanets with masses of $\sim 10^{26}$ g 
and larger form throughout the grid.  When 
(i) the mass in these `oligarchs' is roughly comparable to 
the mass in planetesimals and 
(ii) the surface density in oligarchs exceeds 2--3 g 
cm$^{-2}$ at 1 AU, strong dynamical interactions among oligarchs 
produce a high merger rate which leads to the formation 
of several terrestrial planets.  In disks with lower surface 
density, milder interactions produce several lower mass 
planets.  In all disks, the planet formation timescale is 
$\sim$ 10--100 Myr, similar to estimates derived from the 
cratering record and radiometric data.
\end{abstract}


\section{INTRODUCTION}

In the protosolar nebula, the growth of terrestrial planets 
begins with collisions and mergers of planetesimals, solid 
objects with radii of $\sim$ 1 km 
\citep[e.g.][]{saf69,lis87,ws93,wei97}.  
Initially, planetesimals grow slowly. As they grow, dynamical 
friction circularizes the orbits of the largest objects; 
viscous stirring excites the orbits of the smallest objects.  
Slow, orderly growth ends when the gravitational cross-sections
of the largest objects exceed their geometric cross-sections.
Because dynamical friction is faster than accretion, the 
largest objects stay on circular orbits and grow faster
and faster relative to the smallest planetesimals. This 
runaway growth rapidly concentrates solid material into a 
few protoplanets 
\citep[e.g.,][]{gre78,gre84,wet89,ws93,wei97,kl98,kok00}.

During runaway growth, protoplanets stir the orbits 
of the leftover planetesimals. Stirring reduces 
gravitational focusing factors and slows the growth 
of the protoplanets \citep{ws93,ida93,kb02,raf03c}.  
Although protoplanets grow slowly, they grow faster 
than the leftover planetesimals and intermediate mass 
objects do. These large `oligarchs' slowly clear their
orbits of smaller objects and reach a maximum `isolation mass' 
that depends on the initial surface density of solid material
\citep{lis87,lis93,kok98,raf03c,gol04b}.

Throughout oligarchic growth, protoplanets repel other
oligarchs in the disk \citep{kok96,kok98,kok00,kom02,tho03}. 
Eventually, however, dynamical interactions produce 
collisions and gravitational scattering among the oligarchs 
\citep{cha01,kok02,kom02,tho02,raf03c,gol04b}. During this phase
of `chaotic growth', the largest oligarchs merge with and 
clear away other oligarchs and smaller objects to become 
full-fledged planets \citep[e.g.,][]{gol04a}. 

Within this framework, the transition from oligarchic growth 
to chaotic growth is poorly understood.  Although analytic
estimates provide a guide to the late evolutionary stages 
\citep[e.g.,][]{raf03c,gol04a}, numerical calculations are
necessary to test the basic theory and to derive the end states
and timescales as a function of initial conditions \citep[e.g.][]{kom02}.
Published calculations do not test this regime of the theory 
in much detail.  Pure coagulation and simple hybrid calculations 
cannot follow the transition accurately \citep{ws93,wei97}. 
Direct $n$-body calculations with planetesimals are 
computationally expensive and tend to focus on evolution 
during oligarchic growth, when orbits of individual objects
are easier to track \citep[][and references therein]{ale98,kok02}.
Direct $n$-body calculations without planetesimals follow
dynamical interactions after the transition and concentrate
on the clearing phase \citep{cha01}. Thus, understanding 
this evolutionary phase requires new calculations 
\citep[see the discussion in][]{kok02}.

Here, we use numerical calculations with a new hybrid 
$n$-body--coagulation code to investigate the transition 
from oligarchic growth to chaotic growth.  Our approach 
allows us to combine statistical algorithms for the 
planetesimals with direct $n$-body integrations of the 
oligarchs.  From several simple test cases and complete
planet formation calculations, we show that oligarchic 
growth becomes chaotic when the orbits of oligarchs begin 
to overlap. If the surface density in oligarchs exceeds
a critical value, this transition occurs when the oligarchs
contain roughly half of the mass in solids.  At 1 AU, the 
critical initial surface density is $\Sigma_c \approx$ 
2--3 g cm$^{-2}$.  Thus, oligarchs can make the transition 
from oligarchic to chaotic growth in disks with masses
comparable to the minimum mass solar nebula, where 
$\Sigma \approx$ 8--10 g cm$^{-2}$ at 1 AU.

In disks where the surface density of solids is below the 
limit for chaotic growth, oligarchs slowly merge to form
larger objects. Pairwise interactions, instead of large-scale
chaos, drive the dynamics of these systems. Milder, slower
interactions between oligarchs then produce less massive planets.

We develop the background for our calculations in \S2, describe a
suite of calculations in \S3, and conclude with a brief summary
and conclusions in \S4.

\section{BACKGROUND}

The evolution from planetesimals to planets is marked by
several phases -- orderly growth, runaway growth, oligarchic 
growth, and chaotic growth -- with clear transitions in 
the dynamics and mutual interactions of the ensemble of 
solid objects.  Analytic derivations and sophisticated 
coagulation and $n$-body calculations identify the 
physics of these transitions. Here, we summarize some basic 
results to provide the context for our numerical simulations.

Most considerations of planet formation begin with small objects,
$r_i \lesssim$ 1 km, that contain all of the solid material.
For these sizes, collisional damping and viscous stirring 
roughly balance for orbital eccentricity $e \sim 10^{-5}$. The 
gravitational binding energy, $E_g \sim 10^4$ erg g$^{-1}$, 
is then comparable to the typical collision energy at 1 AU, 
$E_c \sim 10^3$--$10^4$ erg g$^{-1}$. Both energies are
smaller than the disruption energy -- the collision energy 
needed to remove half of the mass from the colliding pair 
of objects -- which is $Q_d \gtrsim 10^7$ erg g$^{-1}$
for rocky material \citep{dav85,ben99}.  Thus, collisions 
produce mergers instead of debris. 

Initially, growth from mergers is slow. The collision 
cross-section is the product of the geometric cross-section
and the gravitational focusing factor, 
\begin{equation}
\sigma_c \sim \pi r^2 f_g \sim \pi r^2 (1 + \beta (e v_K /v_{esc})^2) ~ ,
\end{equation}
where $r$ is the particle radius, $v_K$ is the orbital velocity,
$v_{esc}$ is the escape velocity, and $\beta$ is a coefficient
that accounts for three-dimensional orbits in a rotating disk
\citep{grz90,spa91,ws93}.  Because $e v_K \approx v_{esc}$, 
gravitational focusing factors are small. Thus, growth is 
slow and orderly \citep{saf69}.

As larger objects form, the smaller objects effectively damp 
the orbital eccentricity of larger particles through dynamical 
friction \citep[e.g.][]{ws93,kok95,kl98}.  Viscous stirring 
by the large objects heats up the orbits of the small objects.
In the case where gas drag is negligible, \citet{gol04b} derive 
a simple relation for the ratio of the eccentricities of the 
large (`l') and the small (`s') objects in terms of their 
surface densities $\Sigma_{l,s}$ 
\citep[see also][2003b, 2003c]{kok02,raf03a},
\begin{equation}
\frac{e_l}{e_s} \sim \left ( \frac{\Sigma_l}{\Sigma_s} \right )^n ~ ,
\end{equation}
with $n =$ 1/4 to 1/2.  For $\Sigma_l / \Sigma_s \sim$ $10^{-3}$
to $10^{-2}$, $e_l/e_s \approx$ 0.1--0.25.  Because $e_s v_K 
\lesssim$ $v_{l,esc}$, gravitational focusing factors are large.
Runaway growth begins.

Runaway growth concentrates more and more mass in the largest 
objects.  Dynamical friction produces the largest gravitational
focusing factors among the largest objects. These protoplanets
run away from the smaller objects and contain an ever growing
fraction of the total mass. At the same time, the large objects
continue to stir the leftover planetesimals. The leftovers
have orbital velocity dispersions comparable to the escape 
velocities of the oligarchs. With $e_s v_K \sim v_{esc}$,
equations (1) and (2) show that collision rates decline as 
runaway growth continues. The ensemble of protoplanets and 
leftover planetesimals then enters the oligarchic phase, where 
the largest objects grow faster than the leftover planetesimals.

Among the oligarchs, the smaller oligarchs grow the fastest. 
Each oligarch tries to accrete all of the material in an annular 
`feeding zone' set by balancing the gravity of neighboring
oligarchs.  Within each annulus, each oligarch stirs up the 
remaining planetesimals along its orbit. Because smaller
oligarchs orbit in regions with smaller $\Sigma_l/\Sigma_s$, 
equations (1) and (2) show that smaller oligarchs have larger 
gravitational focusing factors. Thus smaller oligarchs grow 
faster \citep[e.g.,][]{kok98,gol04b}.

During oligarchic growth, protoplanets become isolated from
their surroundings \citep{lis93,kok98,kok00,tho03,gol04a,gol04b}. 
As oligarchs accrete smaller planetesimals, dynamical 
interactions push them apart to maintain a typical separation 
of 5--10 Hill radii, $R_H = (m / 3 m_{\odot})^{1/3}$,
where $m$ is the mass of an oligarch and $m_{\odot}$ is 
the mass of the Sun. The separation depends weakly on the 
mass and semimajor axis of the protoplanet and the local 
surface density \citep{kok98,kok00}.

Oligarchic growth limits the mass of the largest protoplanets
\citep{lis87,lis93,kok98}. If an oligarch accretes all 
planetesimals within a torus of width $a_{acc} = n R_H$, 
it has a mass $m \approx$ 0.005--0.01 $n^{3.2} m_{\oplus}$, 
where $m_{\oplus}$ is the mass of the Earth. For $n \sim$ 5, 
the so-called isolation mass is $m_{iso} \sim$ 0.1 $m_{\oplus}$.
Unless oligarchs move out of their feeding zones or additional 
material is brought into the feeding zone, 
they cannot grow beyond the isolation mass.

A transition from oligarchic growth to chaotic growth appears
to provide the best way for oligarchs to evolve into planets
\citep[e.g.,][]{kom02,gol04b}.
Small oligarchs on roughly circular orbits prevent radial
drift by repelling planetesimals and opening up a gap in
the feeding zone \citep{raf01,raf03a,raf03b,raf03c,raf03d}. 
Larger oligarchs that stir up planetesimals outside the gap 
can grow up to the isolation mass, but no further 
\citep{raf03c,gol04a,gol04b}.
Once the mass in the oligarchs exceeds the mass in leftover
planetesimals, dynamical interactions between the oligarchs 
move them outside their feeding zones. The oligarchs then
compete for the remaining planetesimals 
\citep{gol04a,gol04b}. \citet{gol04a} propose that
large-scale dynamical interactions begin when 
$\Sigma_l \gtrsim \Sigma_s$.  However, this regime has not
been investigated in detail by numerical simulations.

To evaluate the conditions necessary for chaotic orbits to 
produce terrestrial planets, we calculate the formation of 
planets from 1--10 km planetesimals. We start with several 
test calculations to demonstrate and to examine the physical 
processes involved in oligarchic growth and the transition 
from oligarchy to chaos. We then consider the growth of 
planets in a small torus at 0.84--1.16 AU and a 
large torus at 0.4--2 AU.

To provide a measure of the orbital interactions in the
calculations, we define two parameters for the ensemble of 
oligarchs. From published $n$-body calculations, we expect 
significant orbital interactions between oligarchs when their 
typical separations are less than 5--10 $R_H$ 
\citep{kok95,kok98,tho03,bk06}. With $N$ equal mass oligarchs 
spaced evenly in radial distance within a torus of width 
$\Delta a$, oligarchs should begin to interact when 
$5-10 N R_H \approx \Delta a$.  Generalizing this idea to $N$
oligarchs of any mass, we expect significant orbital interactions
among oligarchs when the sum of their Hill radii is $\sim$ 0.1--0.2 
$\Delta a$. To provide a measure of the onset of interactions, 
we define a Hill parameter as the normalized sum of Hill radii 
for all oligarchs:
\begin{equation}
p_H = \frac{\sum_{i=1}^N R_{H,i} }{a_{out} - a_{in}} ~ ,
\end{equation}
where $R_{H,i}$ is the Hill radius for each of $N$ oligarchs
and $a_{in}$ ($a_{out}$) is the inner (outer) radius of the
coagulation grid. We expect significant dynamical interactions
due to orbit overlap when $p_H \gtrsim$ 0.1.

The Hill parameter is useful, because we can relate $p_H$ to the 
surface density required for overlapping orbits. If $m_t$ is the 
typical mass of an oligarch in equation (3), the surface density 
in oligarchs is roughly
\begin{equation}
\Sigma_l \approx 2.5 \left ( \frac{p_H}{0.1} \right ) \left ( 
\frac{m_t}{\rm 10^{26} ~ g} \right )^{2/3} ~ {\rm g ~ cm^{-2}}.
\end{equation}
If oligarchic growth ends when $p_H \approx$ 0.1, equation (3)
yields $\Sigma_l \sim$ 2--3 g cm$^{-2}$, which is 25\% to 33\% of
the mass in a minimum mass solar nebula \citep{wei77,hay81}.

To provide a second measure of the transition from oligarchic to 
chaotic growth, we derive an orbit crossing parameter. We define 
the absolute value of the difference in the semimajor axis of two 
oligarchs, $a_{sep,ij}$ = $| a_i - a_j | $. For each oligarch 
with $j \neq i$, we evaluate $x_{ij} = (a_{sep} - a_j e_j) / R_{H,ij}$, 
where $e_j$ is the orbital eccentricity and 
\begin{equation}
R_{H,ij} = \left ( \frac{m_i + m_j}{3 m_{\odot}} \right )^{1/3} \left ( 
\frac{a_i + a_j}{2} \right )
\end{equation}
is the mutual Hill radius. We find the minimum value
of $x_{ij}$, $x_{min,i} = min(x_{ij})$.
The orbit crossing parameter is then
\begin{equation}
p_o = \frac{\sum_{i=1}^N m_i x_{min,i}}{m_{tot}} ~ ,
\end{equation}
where $m_{tot}$ is the total mass in oligarchs. When 
$p_o \gtrsim$ 5, the orbits of oligarchs do not cross. 
When $p_o \sim$ 0, most orbits cross. We expect significant 
orbital interactions, mergers, and chaotic growth when 
$p_o \lesssim$ 0. 

\section{NUMERICAL CALCULATIONS}

\subsection{Methods}

To explore the end of oligarchic growth, we consider 
numerical calculations of planet formation. 
Because the initial number of planetesimals is large, 
$\sim$ $10^8$ to $10^9$, a statistical approach is the 
only plausible method to derive the collisional growth 
of planetesimals \citep[e.g.][]{saf69,ws93}. When 
a few objects contain most of the mass, the statistical
approach fails. N-body methods can then treat the 
behavior of the largest objects but cannot follow the
evolution of the leftover planetesimals. Here, we use 
a hybrid n-body--coagulation code, which combines the
statistical and n-body approaches to follow the growth 
of 1--10 km planetesimals into planets. 

\citet[][and references therein]{kb04a} summarize our
multi-annulus coagulation code.  The model grid contains 
$N$ concentric annuli with widths $\delta a_i$ centered 
at heliocentric distances $a_i$.  Each annulus contains 
$n(m_{ik},t$) objects of mass $m_{ik}$ with orbital 
eccentricity $e_{ik}(t)$ and inclination $i_{ik}(t)$ in
$M$ mass batches.  Our solution of the coagulation and 
Fokker-Planck equations treats elastic and inelastic
collisions between all particles in all annuli.  We adopt 
collision rates from kinetic theory -- the particle-in-a-box 
method -- and use an energy-scaling algorithm to assign 
collision outcomes.  We derive changes in orbital parameters 
from gas drag, dynamical friction, and viscous stirring 
\citep{ada76, oht02}.  

\citet{bk06} describe the n-body code and our methods for 
combining n-body calculations with the coagulation code.
When an object in the coagulation code reaches a preset mass,
it is `promoted' into the n-body code. To set the initial
orbit for this object, we use the three coagulation coordinates, 
$a$, $e$, and $i$, and select random values for the longitude
of periastron and the argument of perihelion. Because the
annuli have finite width $\delta a_i$, where $i$ is the index,
we set the semimajor axis of the promoted object to $a_p$ = 
$a_i + (0.5 - x) \delta a_i$, where $x$ is a random number between 
0 and 1.  When two or more objects within an annulus are promoted 
to the n-body code during the same timestep, we restrict the 
choices of the orbital elements to minimize orbital interactions
between the newly promoted $n$-bodies. 

We follow the mutual interactions of the ensemble of `n-bodies', 
including mergers, using a robust set of integrators. These 
integrators include accretion and drag terms that link the 
evolution of the planetesimals to the evolution of the n-bodies. 
To compute how rapidly n-bodies accrete planetesimals, we adopt 
a particle-in-a-box formalism. Direct calculations for gas drag
and Fokker-Planck gravitational interactions provide rates for
changes in $a$, $e$, and $i$ for each n-body due to interactions
with planetesimals.

\cite{bk06} describe tests of the hybrid code along with initial
results for terrestrial planet formation. When the Hill radius 
of a promoted n-body is small compared to the separation of
annuli in the coagulation grid, the hybrid code reproduces 
published results of previous investigations
\citep[e.g.][]{grz90,ws93,wei97,dun98,cha01}. 
In the calculations described below, we set the promotion 
mass to $m_{pro} =$ 1--30 $\times 10^{24}$ g, which yields
a small Hill radius, $\sim$ 0.0005--0.002 $a$, compared to the 
separation of adjacent annuli in the coagulation grid, $\sim$ 
0.01--0.04 $a$.

Our numerical calculations begin with a mass distribution of 
planetesimals in 32--64 concentric annuli with initial surface 
density $\Sigma = \Sigma_0$ (a/1 AU)$^n$, with $n$ = 1--1.5. 
The spacing between 
successive mass batches is $\delta = m_{i+1}/m_i$.  We adopt 
$\delta$ = 1.4--2.0 for most calculations.  All planetesimals 
start with initial eccentricity $e_0$ and inclination $i_0$. 
As planetesimals evolve, the average mass $m_{ik}$ and orbital 
parameters $e_{ik}$ and $i_{ik}$ of each mass batch change. 
We add mass batches as planetesimals grow in mass and reserve 
8--16 mass batches in each annulus for n-bodies.  Throughout 
the calculation, gas drag and collisions transport planetesimals 
in the coagulation code from one annulus to another.  In addition 
to these processes, mutual gravitational interactions can scatter 
n-bodies into different annuli.

For these calculations, we assume a simple treatment of the gas.
The midplane density is
\begin{equation}
\rho_g(a,t) = \rho_0(a) e^{-t/t_0},
\end{equation}
where $\rho_0(a)$ is the initial gas density and $t_0$ is a constant.
We adopt a gas surface density $\Sigma_g = 100 \Sigma$ and set 
$\rho_0 = \Sigma_g / 2H$, where $H$ is the gas scale height \citep{kh87}.
Consistent with observations of pre-main sequence stars, we adopt
$t_0$ = 1 Myr \citep[see, for example,][]{hai01,you04,cal05,dal05}.

\subsection{Simple model}

To isolate the important physical parameters in the evolution
of oligarchs, we first consider an ensemble of equal mass 
planetesimals in 32 concentric annuli at 0.84--1.16 AU. The 
planetesimals have mass $m_s$, surface density $\Sigma$ = 
$\Sigma_s (a / {\rm 1 ~ AU})^{-3/2}$, initial eccentricity 
$e_0$ = $10^{-5}$ and inclination $i_0$ = $e_0/2$. 
Within this swarm, we embed $N$ oligarchs with mass $m_l$
and surface density $\Sigma$ = $\Sigma_l (a / {\rm 1 ~ AU})^{-3/2}$. 
The oligarchs have the same initial eccentricity and 
inclination as the planetesimals.

To evolve this ensemble in time, we calculate gravitational
stirring for all interactions. We allow oligarchs -- but
not planetesimals -- to collide and merge. This assumption 
allows us to focus on the dynamical evolution of the oligarchs 
in the absence of planetesimal accretion. 

Figure 1 shows the evolution of the eccentricity for three sets 
of oligarchs with $\Sigma_l/\Sigma_s$ = 0.5 and $\Sigma_l$ =
1--4 g cm$^{-2}$. The planetesimals have $m_s$ = $10^{24}$ g; 
the oligarchs have $m_l = 10^{26}$ g, comparable to the isolation 
mass for this grid. The Hill parameter increases from 
$p_H = 0.04$ (lower panel) to $p_H$ = 0.07 (middle panel)
to $p_H$ = 0.13 (upper panel).  In each frame, the colored 
tracks indicate the eccentricities of oligarchs in the grid. 
Along the track of a single oligarch, the color changes when 
two oligarchs merge. Although a single color does not track 
the motion of an individual oligarch throughout the evolution,
the ensemble of curves tracks the merger history of the
final set of oligarchs. Tracking backwards in time along connected 
curves yields the evolution of one of the oligarchs that remains
at the end of the calculation.
The legends list $\Sigma_l$ and the initial and final number 
of oligarchs (Figure 2).

The eccentricity evolution is sensitive to the initial mass in
the swarm. For $\Sigma_l$ = 1 g cm$^{-2}$, the oligarchs
have a typical separation of 15--20 $R_H$ and do not have
significant interactions. As the oligarchs stir up the
planetesimals, dynamical friction maintains a constant ratio
$e_s/e_l \sim$ $(m_l/m_s)^{1/2}$ $\sim$ 0.1. Although the
orbits of the oligarchs become more and more eccentric,
the growth of $e_l$ is slow (Figure 1, lower panel). It takes
$\sim$ 1 Myr to reach $e_l \sim$ 0.01 and another 8--9 Myr 
to reach $e_l \sim$ 0.02. After $\sim$ 100 Myr, when $e \sim$
0.04, the orbits begin to overlap.

As $\Sigma_l$ increases, orbital interactions occur on shorter
timescales (Figure 1, middle and upper panels). For $\Sigma_l$
= 2 g cm$^{-2}$, the initial separation of the oligarchs is
$\sim$ 10--12 $R_H$. It takes only $5 \times 10^4$ yr for the
typical eccentricity to reach $e \sim$ 0.01, when the minimum
separation between oligarchs is only $\sim$ 8 $R_H$. At
$\sim 10^5$ yr, $e \sim$ 0.02 and orbits overlap. Two oligarchs
merge at $\sim 2 \times 10^5$ yr; two more merge at $\sim$ 1 Myr.
When we stop the calculation at 1 Myr, two oligarchs remain 
in eccentric orbits, $e \gtrsim$ 0.025 and are likely to merge 
with other oligarchs.

For $\Sigma_l$ = 4 g cm$^{-2}$, orbits begin to overlap in 
$\sim 10^4$ yr. After a single merger at $\sim 10^4$ yr,
orbits continue to grow more eccentric. At $5 \times 10^4$ yr,
all orbits overlap and the merger rate accelerates. There are
two additional mergers at $\sim 10^5$ yr, another two by 
$3 \times 10^5$ yr, and three more by $\sim$ 1 Myr. At 1 Myr, 
all of the remaining oligarchs have eccentric, overlapping
orbits and many are likely to merge over the next few Myr.

Figure 2 illustrates the chaotic behavior of the semimajor 
axis in these test cases. For $\Sigma_l$ = 1 g cm$^{-2}$,
the oligarchs have a constant semimajor axis for almost 100 Myr.
When the total mass is a factor of two larger ($\Sigma_l$
= 2 g cm$^{-2}$), the semimajor axes are constant for $\sim$
$10^5$ yr. Once the orbits start to overlap, several oligarchs
show considerable excursions in semimajor axis of 0.1 AU or
more, $\sim$ 30\% to 40\% of the grid. Two of these oligarchs
merge with other oligarchs. For $\Sigma_l$ = 4 g cm$^{-2}$,
the orbits are very chaotic, with larger radial excursions
and many mergers. 

Figures 3 and 4 show the evolution of the number of 
oligarchs $N_o$ and the orbit crossing parameter $p_o$ 
for the three cases in Figures 1--2 and a fourth case 
with $\Sigma_l$ = 8 g cm$^{-2}$.  Mergers are prominent 
in all but the lower surface density test.
In the tests where mergers are important, $p_o \lesssim 0$
when orbits overlap and mergers begin.

These tests confirm our general expectations. Oligarchs
stir up planetesimals along their orbits. Dynamical friction
maintains a fixed ratio of $e_l/e_s$ (equation 2). Because the 
dynamical friction timescale depends on the surface density, the 
orbits of oligarchs in more massive disks overlap on shorter 
timescales than the orbits of oligarchs in less massive disks. 
More massive disks contain more oligarchs, leading to more
chaotic orbits on shorter timescales.

To provide additional tests, we consider two more sets of 
calculations. Models with fixed $\Sigma_l$ and variable
$\Sigma_s$ gauge the importance of damping by leftover
planetesimals. We first consider models with $m_s = 10^{24}$ g
and then examine calculations with $m_s = 10^{16}$ g.

Figure 5 shows the evolution of semimajor axis for three
calculations with 
$\Sigma_s$ = 2 g cm$^{-2}$ (upper panel), 
$\Sigma_s$ = 4 g cm$^{-2}$ (middle panel), and
$\Sigma_s$ = 8 g cm$^{-2}$ (lower panel). 
The tracks in the upper panel repeat those from the middle
panel of Figure 2; chaos begins at $\sim 10^5$ yr when
the orbits start to overlap. As $\Sigma_s$ increases,
dynamical friction between the small and large objects 
is more efficient. The orbits overlap and chaos begins 
later when $\Sigma_l / \Sigma_s \lesssim$ 0.5. The middle 
set of tracks in Figure 5 demonstrates this behavior:
there is no chaos until the system has evolved for 1 Myr.  
However, in models where $\Sigma_s$ 
exceeds $\sim$ 8 g cm$^{-2}$, viscous stirring among
the planetesimals increases their velocity dispersions
on shorter timescales. Larger viscous stirring results in
larger orbital eccentricities for the oligarchs (equation 2)
and a faster onset of chaos. For $\Sigma_s$ = 8 g cm$^{-2}$,
orbits begin to overlap at $\sim 3 \times 10^5$ yr. These
chaotic interactions grow with $\Sigma_s$.

To test the importance of viscous stirring among planetesimals, 
Figure 6 repeats the calculations of Figure 5 for planetesimals
with $m_s = 10^{16}$ g. In this test, the ratio of orbital
eccentricity is much larger, $e_s/e_l \sim$ 75--80. 
The viscous stirring timescale for the planetesimals is 
much larger than 1 Myr. For all
$\Sigma_s$ = 4--32 g cm$^{-2}$, planetesimal stirring is 
negligible. Because dynamical friction is important, the 
oligarchs remain well-separated and never develop overlapping
orbits.

These tests demonstrate the main physical processes involved
in the transition from oligarchy to chaos. Dynamical friction
maintains a fixed ratio of $e_l/e_s$ (equation 2). Viscous
stirring increases the orbital eccentricities of planetesimals
until the orbits of oligarchs interact.  When large planetesimals
contain a significant amount of mass, they contribute to the
stirring. Otherwise, oligarchs provide all of the stirring.
Once orbits overlap, chaos ensues. Chaos produces mergers and
large excursions in semimajor axis, which starts the process 
that clears out the disk to produce large planets. 

Three parameters -- $\Sigma_l / \Sigma_s$, $p_H$, and $p_o$ -- 
provide good measures of the transition from oligarchy to chaos. 
The Hill parameter measures when the oligarchs have enough mass 
to interact dynamically. The ratio $\Sigma_l / \Sigma_s$ isolates
the time when planetesimals cannot damp the oligarchs and thus
prevent large-scale dynamical interactions. The orbit overlap 
parameter distinguishes times when orbit overlap is important.

To understand the transition from oligarchy to chaos in less
idealized situations, we now consider complete planet formation
simulations using the full hybrid code. The calculations start 
with 1--10 km planetesimals and allow all objects to collide, 
merge, and interact gravitationally.  When objects in the 
coagulation code reach 
$m \approx 2 \times 10^{25}~(\Sigma_0 / {\rm 8~g~cm^{-2}})$ g, 
we promote them into the $n$-body code and follow their individual
trajectories.  We describe calculations in a small (large)
torus in \S3.3 (\S3.4).

\subsection{Planet formation at 0.86--1.14 AU}

The calculations begin with 1--3 km planetesimals in a
torus extending from 0.86 AU to 1.14 AU. We divide this
region into 32 annuli and seed each annulus with planetesimals
in nearly circular and coplanar orbits ($e_0 = 10^{-5}$ and 
$i_0 = e_0/2$). The planetesimals have surface density
$\Sigma = \Sigma_0 (a / {\rm 1 ~ AU})^{-3/2}$, with $\Sigma_0$
= 1--16 g cm$^{-2}$ at 1 AU. In these calculations, we do not
consider fragmentation, which generally speeds up the growth
of the largest objects at the expense of mass loss from
disruptions and gas drag \citep{ws93,kl98}. \citet{wei97}
consider a similar suite of calculations. Where it is
possible to compare, our results agree with these calculations
\citep[see also][]{kom02}.

For $\Sigma_0$ = 8 g cm$^{-2}$, growth at 1 AU follows a
standard pattern \citep{ws93, wei97, kl98}. After a few 
thousand years, mergers produce a few large objects with 
radii of $\sim$ 10 km. As dynamical friction circularizes
the orbits of these objects, runaway growth begins. It
takes only $10^4$ yr to produce several dozen 100--300 km
objects. At $\sim 2 \times 10^4$ yr, the first object is
promoted into the $n$-body code. As larger objects form
farther out in the disk, more promotions occur. These
objects continue to grow rapidly until they reach `isolation'
masses of $\sim 10^{26}$ g, when stirring begins to reduce 
gravitational focusing factors. 

The transition to oligarchic growth begins at the inner 
edge of the grid and rapidly propagates outwards. At $\sim$ 
$3 \times 10^5$ yr, the number of oligarchs with masses 
$m \gtrsim$ $10^{26}$ g peaks at $N_o \sim$ 7. Soon after 
oligarchic growth begins at the outer edge of the grid, 
oligarchs at the inner edge of the grid begin to interact 
dynamically. A wave of strong dynamical interactions then 
moves out through the grid. It takes $\sim$ 1 Myr for the 
wave to move from $\sim$ 0.85 AU to $\sim$ 1.15 AU. During 
this period, some oligarchs merge. Others migrate through 
the grid on highly eccentricity orbits. 

From $\sim$ 1 Myr onward, mergers slowly reduce $N_o$.
It takes $\sim$ 1 Myr for the first 2 mergers and another 
$\sim$ 2 Myr for the second 2 mergers. After 100 Myr, 
only 3 oligarchs remain. One of these has $m \sim$ 0.43 
$m_{\oplus}$, $a \sim$ 0.9 AU, and $e \sim$ 0.08. 
The other two oligarchs have $m \sim$ 0.05 $m_{\oplus}$
and $e \sim$ 0.1 (Figure 7).  Aside from the eccentricity 
of the more massive planet, the properties of these objects 
are reasonably close to those of the Earth and Mars. 
Fragmentation and interactions with the gas probably promote 
smaller eccentricities for the largest objects 
\citep[e.g.,][]{ws93,agn02,kom02}.

Figure 7 illustrates the evolution of the semimajor axes
of the oligarchs for three calculations with 
$\Sigma_0$ = 2 g cm$^{-2}$ (lower panel),
$\Sigma_0$ = 4 g cm$^{-2}$ (middle panel), and
$\Sigma_0$ = 8 g cm$^{-2}$ (upper panel). 
As in Figure 1, the tracks change color when two oligarchs
collide and merge. The labels indicate the final mass, 
in Earth masses, of the largest oligarchs at 100 Myr.

The timescale for the onset of chaotic growth depends on 
the initial surface density. More massive disks reach the 
transition first.
\citep[see, for example,][]{lis87}. For $\Sigma_0$
= 8 g cm$^{-2}$, the transition begins at $\sim$ a few
$\times ~ 10^5$ yr. The transition is delayed to $\sim$
1 Myr for $\Sigma_0$ = 2 g cm$^{-2}$. 

The character of the transition to chaotic growth also 
depends on the initial surface density. In relatively
massive disks with $\Sigma_0$ $\sim$ 8 g cm$^{-2}$, many 
oligarchs develop highly eccentric orbits and exhibit
large variations in their semimajor axes. These large
excursions result in many mergers and a rapid reduction
in $N_o$. In less massive disks with
$\Sigma_0 \lesssim$ 2--4 g cm$^{-2}$, only 1 or 2 oligarchs
develop highly eccentric orbits. Most mergers are caused
by two-body interactions, instead of large-scale dynamical
interactions throughout the grid.

Figures 8--10 illustrate these general conclusions. 
In Figure 8, the orbit crossing parameter rapidly
approaches zero for calculations with $\Sigma_0$ = 
8 g cm$^{-2}$. At 0.1--1 Myr, $p_o$ has a long plateau;
close approaches between oligarchs cause $p_o$ to fall 
below zero; mergers cause $p_o$ to jump above zero. 
During a series of 4 mergers at 10 Myr, $p_o$ remains 
below 0 for a long period.  After the final merger, 
$p_o$ jumps to 40, where it remains for many Myr.
For smaller $\Sigma_0$, $p_o$ remains well above zero until
one or two close pairwise interactions pushes it below zero.
Once these interactions produce a merger, the
systems stabilize and $p_o$ moves well above zero.

Figure 9 shows the time evolution of the total number of
$n$-bodies within the grid.  During the transition from 
runaway growth to oligarchic growth, $N_o$ peaks and 
remains constant for 0.1--1 Myr.  Once orbital interactions 
begin, $N_o$ declines.  The rate of decline is fastest in 
the most massive systems.

The Hill parameter also shows the rapid transition from
oligarchy to chaos. When $N_o$ peaks, $p_H$ reaches a maximum. 
It declines sharply when mergers reduce the number of oligarchs. 
After a merger or series of mergers, $p_H$ rises slowly as 
the remaining oligarchs accrete leftover planetesimals.

To examine the sensitivity of these results to $m_{pro}$,
we consider a broad range of promotion mass.  For 
$\Sigma_0$ = 1--16 g cm$^{-2}$ at 0.8--1.2 AU, calculations 
with $m_{pro} \gtrsim$ $2 \times 10^{25}$ 
($\Sigma_0 / {\rm 4~g~cm^{-2}}$) g cannot follow the evolution 
of the most massive coagulation particles accurately. Thus, 
the production rate of oligarchs is inconsistent.  Once
a few oligarchs form, the models provide a poor treatment
of interactions among oligarchs and therefore yield poor
estimates of the merger rate and the timescale for the transition 
from oligarchy to chaos.

Calculations with smaller $m_{pro}$ provide better solutions
to the evolution of the largest objects.  To measure the 
quality of the results for the transition from oligarchy to 
chaos, we define $t_{\Sigma}$ as the time when oligarchs with 
$m \gtrsim 10^{26} (\Sigma_0 / {\rm 8~g~cm^{-2}})$ g contain 
half of the initial mass and $t_m$ as the time when these 
oligarchs start to merge and the number of oligarchs starts 
to decline. The scaling of the oligarch mass with $\Sigma_0$ 
provides a clean way to compare models with different initial 
conditions.

Our calculations show clear trends for $t_m$ and $t_{\Sigma}$ 
as a function of $\Sigma_0$ (Figure 11). The timescale for 
oligarchs to contain half of the initial mass roughly varies as 
$t_{\Sigma} \propto \Sigma_0^{-2/3}$ with very little 
dispersion\footnote{The origin of this
relation lies in the growth and stirring rates. In the absence
of stirring, the growth time is $t \propto \Sigma_0^{-1}$. Because
more massive oligarchs form in more massive disks, the vertical
scale height $H$ of leftover planetesimals is larger in more 
massive disks. Larger scale heights reduce the growth rate by 
roughly $H^{1/3} \propto \Sigma_0^{1/3}$, which results in
$t_{\Sigma} \propto \Sigma_0^{-2/3}$.}. The transition from oligarchic 
growth to chaotic grwoth at $t = t_m$ also depends on $\Sigma_0$.  Our 
results yield an approximate relation, $t_m \propto \Sigma_0^{-3/2}$. 
However, this trend appears to have inflection points for 
$\Sigma \lesssim$ 2 g cm$^{-2}$ and $\Sigma \gtrsim$ 8 g cm$^{-2}$.  
Calculations in progress will allow us to test this relation and 
its origin in more detail.

There are no large trends for $t_m$ and $t_{\Sigma}$ with $m_{pro}$. 
Although calculations with smaller $m_{pro}$ yield smaller dispersions 
in $t_m$ and $t_{\Sigma}$ at each $\Sigma_0$, the median values for 
$t_m$ and $t_{\Sigma}$ are fairly independent of $m_{pro}$. A 
Spearman rank test \citep{pre92} suggests a weak inverse correlation 
between $t_m$, $t_{\Sigma}$ and $m_{pro}$, which we plan to test with 
additional calculations. 

Independent of $m_{pro}$, our results demonstrate a clear trend of 
the ratio $r_{\Sigma} = t_m/t_{\Sigma}$ with initial surface density 
(Table 1).  In massive disks with $\Sigma_0 \gtrsim$ 12 g cm$^{-2}$, 
mergers among oligarchs begin before oligarchs contain half of the 
mass ($t_m/t_{\Sigma} \lesssim 1$). In low mass disks with $\Sigma_0$
$\lesssim$ 2 g cm$^{-2}$, oligarchs start to merge after they contain 
half of the mass ($t_m/t_{\Sigma} \gtrsim 1$). 

To test whether variations in the frequency and strength of dynamical 
interactions among oligarchs cause the trends in $t_m$ and $r_{\Sigma}$, 
we define a `nearest neighbor parameter' $n_n$ which measures the 
average number of oligarchs within 10 $R_H$ of another oligarch. 
From \S3.2, configurations with $n_n \gtrsim$ 1 lead to strong 
dynamical interactions among the oligarchs; oligarchs interact mildly 
when $n_n \lesssim$ 1. Table 1 lists the average of the maximum 
value of $n_n$ and its dispersion for our calculations. 
Disks with $r_{\Sigma} \gtrsim$ 0.5 have $n_{n,max} \lesssim$ 1;
disks with $r_{\Sigma} \lesssim$ 0.5 have $n_{n,max} \gtrsim$ 1.
This result confirms the visual impressions from Figures 7--10:
dynamical interactions among oligarchs are stronger in massive
disks and milder in low mass disks.

To conclude this section, Figure 12 shows the evolution of the 
mass for each oligarch in one calculation. Here, points of one 
color correspond to the track of one oligarch. From $\sim 10^4$ yr 
to $\sim 10^5$ yr, large objects form and grow rapidly.  Once 
stirring reduces gravitational focusing factors, growth slows.  
During the late stages of runaway growth and the early stages 
of oligarchic growth, several neighboring oligarchs merge. As 
their gravitational reach extends, these larger oligarchs grow 
faster and faster.  Smaller oligarchs cannot compete for leftover 
planetesimals and grow slowly. 

During chaotic growth, the largest oligarchs merge with smaller
oligarchs. As the small oligarchs are depleted, the merger
rate slows. With highly eccentric orbits, a few small oligarchs
last for 10--30 Myr before colliding with a large oligarch. 
After 100 Myr, all but one small oligarch have collided and
merged with the two large planets that remain at the end of 
the evolution.

\subsection{Planet formation at 0.4--2 AU}

These calculations begin with 5 km planetesimals in a
torus extending from 0.4 AU to 2 AU. We divide this
region into 40 annuli and seed each annulus with planetesimals
in nearly circular and coplanar orbits ($e_0 = 10^{-5}$ and 
$i_0 = e_0/2$). To provide a contrast with previous simulations,
the planetesimals have surface density 
$\Sigma = \Sigma_0 (a / {\rm 1 ~ AU})^{-1}$, with 
$\Sigma_0$ = 2--16 g cm$^{-2}$ at 1 AU. Calculations 
with the more standard $\Sigma \propto r^{-3/2}$ yield 
similar results.  As in \S3.3, we do not consider 
fragmentation. This torus is larger than the 0.5--1.5 AU
region examined by \citet[][see also Kominami \& Ida 2002]{wei97},
but similar to the torus described by \citet{cha01}. 
\citet{cha01} starts
his calculations with lunar mass objects, instead of
planetesimals. \citet{bk06} compare results derived from 
our hybrid code for calculations starting with 1--10 km
planetesimals or 100--200 lunar mass objects. When we
start with smaller objects, our calculations produce fewer 
oligarchs on shorter timescales than the \citet{cha01}
calculations.

Growth in the larger torus is similar to that in the smaller 
torus.  For $\Sigma_0$ = 8 g cm$^{-2}$, mergers produce a 
few large objects with radii of $\sim$ 10 km at 0.4 AU in 
roughly a thousand years. Once runaway growth begins, it
takes only $\sim 10^4$ yr for the first promotion into the 
$n$-body code. As runaway growth propagates outward in 
heliocentric distance, objects throughout the grid reach the 
isolation mass of $\sim 10^{26}$ g. It takes less than $10^5$ 
yr to produce 10 isolated objects and another $2 \times 10^5$ 
yr to produce the next 10 isolated objects.  These objects 
continue to grow and reach typical masses of 0.05--0.1 
$m_{\oplus}$ in $\sim$ 1 Myr.

During oligarchic growth, occasional two body interactions
produce mergers of oligarchs. These mergers always occur
when the coagulation code produces 2--3 oligarchs in the 
same annulus. Often, dynamical interactions move one or
two oligarchs into neighboring annuli, where they begin to 
accrete planetesimals more rapidly. Occasionally, dynamical
interactions between several oligarchs in neighboring annuli 
produce one or two mergers, which stabilizes the system 
locally and leads to the isolation of the remaining 2 or
3 oligarchs.

As oligarchs start to form at the outer edge of the grid,
oligarchs at the inner edge of the grid begin to interact
dynamically (Figure 12). For $\Sigma_0$ = 8 g cm$^{-2}$,
these interactions start at $\sim$ 1 Myr. Chaotic interactions
then spread throughout the grid.  For the next $\sim$ 10 Myr,
mergers produce fewer but larger oligarchs, which rapidly
sweep up the remaining planetesimals. After $\sim$ 200 Myr,
only 5 oligarchs remain. The largest have masses comparable 
to those of the Earth and Venus. Smaller oligarchs have
masses comparable to the mass of Mars.

As in \S3.3, the evolution in less massive disks proceeds more 
slowly and less chaotically. The magnitude of dynamical orbital 
interactions is non-linear: once chaos starts in a massive disk, 
it rapidly propagates throughout the grid. Pairwise interactions 
dominate the dynamics of lower mass disks. These interactions 
usually do not affect other oligarchs.

Figures 13--14 show the time evolution of $p_o$ and $p_H$. 
In all calculations, the typical orbital separation of 
the largest objects rapidly approaches zero when the 
first oligarchs form (Figure 13). This transition occurs 
sooner in more massive disks. During oligarchic growth, 
the orbits of oligarchs slowly move closer and closer.
The number of oligarchs and the average Hill radius of 
the oligarchs (Figure 14) rise dramatically; these peak 
when $\Sigma_l / \Sigma_s$ $\sim$ 0.4--0.5. Once 
$\Sigma_l / \Sigma_s \gtrsim$ 0.5, chaotic interactions 
cause oligarchs to merge. When the orbits overlap,
large-scale chaos leads to rapid mergers and planets with
masses comparable to the mass of the Earth. In less massive
disks with $\Sigma_0 \lesssim$ 2 g cm$^{-2}$, the oligarchs
remain well-separated relative to their Hill radii (Figure 14).
Pairwise interactions between oligarchs then produce most of 
the mergers. This evolution is slow and results in more 
planets with masses comparable to the mass of Mercury.

Finally, Figure 15 shows the growth of the oligarchs for
a calculation with $\Sigma_0$ = 8 g cm$^{-2}$. Here, tracks
of a single color show the time evolution in the mass of a
single oligarch. Tracks end when an oligarch merges with 
another oligarch in the grid.

The tracks in Figure 15 clearly illustrate the three stages 
of growth in the terrestrial zone. Most tracks are initially
steep and then slowly turn over.  The dramatic changes in the
slope mark the transition from runaway growth to oligarchic
growth. At $\sim 10^5$ yr, this transition propagates as a 
wave from the inner part of the grid at 0.4 AU to the outer
part of the grid at 2 AU.  

The onset of distinct steps in mass indicates the transition
from oligarchic growth to chaotic growth. Starting at $\sim$
1--3 Myr, this transition tales $\sim$ 3--10 Myr to propagate 
from 0.4 AU to 2 AU. Once the full grid is chaotic, mergers
rapidly separate the oligarchs into small oligarchs, with
$m \sim$ 0.01--0.1 $m_{\oplus}$ and $e \gtrsim$ 0.1, and 
large oligarchs, with $m \gtrsim$ 0.2--1 $m_{\oplus}$ and 
$e \lesssim$ 0.05. After $\sim$ 100 Myr, this division is
stark: there are 3 oligarchs with $m \gtrsim 0.5 ~ m_{\oplus}$
and 3 others with $m \lesssim 0.07 ~ m_{\oplus}$. 

Calculations with a large range in $m_{pro}$ provide a measure
of the quality of these conclusions. Because the 0.4--2 AU grid 
is much larger than the 0.84--1.16 AU grid, these calculations
produce more oligarchs and are less sensitive to $m_{pro}$ than
calculations in a smaller grid. Nevertheless, calculations with
$m_{pro} \gtrsim$ $3 \times 10^{25}$ 
($\Sigma_0 / {\rm 4~g~cm^{-2}}$) g yield a much larger range in
the evolution timescales than calculations with smaller $m_{pro}$.
As in \S3.3, models with $m_{pro} \sim$ $1-10 \times 10^{24}$ g
yield consistent results for these timescales.

\subsection{Other Physics}

In this set of calculations with the hybrid code, we concentrate 
on mergers and dynamical interactions between solid objects. We 
ignore fragmentation and interactions with the gas. In previous
calculations, fragmentation of 1--10 km planetesimals produces 
cm- to m-sized fragments which are more closely coupled to the 
gas than larger planetesimals. Gas drag circularizes the 
orbits of these objects, which can then be accreted more 
rapidly than the leftover planetesimals. From previous numerical 
calculations, the onset of runaway and oligarchic growth are 
$\sim$ 25\% sooner than illustrated in \S3.3 and \S3.4
\citep[e.g.,][]{ws93,kl99,kb04b,kb05}.

Fragmentation is important for setting the visibility of the disk
\citep{kb04b}.  The rate of debris production from fragmentation
typically peaks during the transition from runaway to oligarchic
growth.  If fragmentation produces an efficient collisional cascade, 
radiation pressure and Poynting-Robertson drag may eject small
particles before the oligarchs can accrete the fragments. The disk
then produces a substantial infrared excess \citep{kb04b}.
If some process halts or slows the cascade, oligarchs can accrete 
the fragments efficiently \citep[e.g.,][]{gol04b}. The disk may 
then produce a modest infrared excess. We have started calculations
to evaluate the importance of fragmentation in setting the
timescales for runaway and oligarchic growth and to estimate
the amount of dusty debris as a function of time.

Interactions with the gas can produce significant evolution in
the eccentricity and orbital semimajor axis of oligarchs 
\citep[e.g.,][]{art93,agn02,kom02,tan02,tan04}. 
In the calculations here, simple gas drag as in \citet{ada76} 
has a negligible impact on the evolution of planetesimals and 
oligarchs. However, interactions with density waves can 
circularize the orbits of oligarchs on short timescales
\citep[e.g.,][]{art93,agn02,kom02,tan04}
and can cause significant inward migration of the orbit
\citep[e.g.][]{art93,tan02}. 
For a gaseous disk with the surface density of a minimum mass 
solar nebula, the circularization timescale for an oligarch 
is $\tau_c \sim$ 0.1--0.2 Myr at 1 AU \citep{agn02,tan04}.  
Significant orbital migration of an oligarch can occur over 
$\tau_m \sim$ 1--3 Myr \citep{tan02,tan04}. 
Thus, interactions with the gas occur on timescales comparable
with timescales for oligarchs to grow and interact dynamically.

The estimates for orbital migration and eccentricity damping
assume a large gas density in the disk. Observations of young 
stars suggest the dust -- and presumably the gas -- disappears
in $\sim$ 1--10 Myr \citep[e.g.,][and references therein]{cal05,dal05}.  
If the gas density declines exponentially with time as in 
equation (7) with $t_0 \sim$ 1 Myr, radial migration may have
little impact on the evolution of oligarchs. Eccentricity
damping, however, may play an important role in the transition
from oligarchic to chaotic growth. We have begun calculations
to see how eccentricity damping and radial migration change 
the outcomes of our calculations. 

Together, fragmentation and interactions with the gas 
probably set the eccentricity of the final ensemble of
planets. If fragmentation is efficient at converting 
intermediate mass objects into smaller fragments, the
fragments can efficiently circularize the orbits of 
growing oligarchs (equation 2; Figures 5--6). Interactions
with the gas also circularizes the orbits of the most massive 
objects. We have started a set of calculations to test
the relative efficiency of damping and fragmentation in
setting the eccentricity of surviving oligarchs.

\section{SUMMARY}

Our calculations with the hybrid $n$-body--coagulation code 
are the first to evolve an ensemble of planetesimals into a
system of a few planets. The calculations reproduce both the 
standard results of coagulation calculations for the early 
stages of planet formation and the results of $n$-body 
calculations for the late stages of planet formation 
\citep[see also][]{bk06}.  In particular, we follow the 
evolution through the critical transition from oligarchic
growth to chaotic growth, where objects evolve from isolated
oligarchs into full-fledged planets.  These calculations 
provide some new insights into the transition from oligarchic 
growth to chaotic growth \citep[see also][]{kok02,kom02}.

In a disk with initial surface density $\Sigma_0 \gtrsim$ 
3--16 g cm$^{-2}$ at 1 AU, collisions and mergers of 
1--10 km planetesimals naturally lead to the formation 
of terrestrial planets with masses ranging from 
0.05--2 $m_{\oplus}$ 
\citep[see also][and references therein]{ws93,wei97,cha01,kok02}.
The growth follows a standard pattern. Slow orderly growth
rapidly gives way to runaway growth, which concentrates much
of the initial mass into many protoplanets with masses $\sim$
$10^{26}$ g.  Stirring of leftover planetesimals by the largest
objects reduces gravitational focusing factors and slows growth.
During oligarchic growth, large oligarchs become isolated and 
slowly accrete the leftovers.

Several factors produce a transition from oligarchic growth to 
chaotic growth \citep[see also][]{kom02}. As the oligarchs grow, 
they contain an ever increasing fraction of the total mass. 
A few oligarchs merge but most remain isolated from other 
oligarchs. As their Hill radii grow, their orbits begin to overlap.  
When the surface density in oligarchs reaches a critical value, 
oligarchs interact chaotically.  As gravitational interactions 
scatter oligarchs throughout the disk, the merger rate increases 
dramatically.  Eventually only a few oligarchs remain in roughly 
circular orbits. 

Our results isolate the two conditions necessary for the transition 
from oligarchy to chaos. When oligarchs contain roughly half of the 
mass of solid material in the disk, $\Sigma_l \sim \Sigma_s$,
dynamical interactions between oligarchs are more important than
dynamical friction from planetesimals \citep[e.g.,][]{gol04b}. 
When the surface density in oligarchs exceeds a critical value, 
$\Sigma_c \approx$ 2--3 g cm$^{-2}$, oligarchs begin to interact
chaotically (\S3).  More massive disks can reach this limit when
$\Sigma_l < \Sigma_s$ (Table 1). In less massive disks, milder 
dynamical interactions begin when $\Sigma_l \gtrsim \Sigma_s$ 
(Figure 11).  If the surface density in oligarchs remains below 
the critical value, interactions between oligarchs are less chaotic 
even when $\Sigma_l \gtrsim \Sigma_s$.  Interactions among 2 or 3 
oligarchs produce a small merger rate which eventually yields 
a system with lower mass planets compared to more massive disks.

Although dynamical interactions among the ensemble of oligarchs 
produce terrestrial mass planets in all disks, more massive disks 
yield more massive planets. Our results suggest a maximum mass,
$m_{max} \sim$ 1--2 $m_{\oplus}$ for $\Sigma_0 \sim$ 8--16 g cm$^{-2}$
and $m_{max} \sim$ 0.1--0.2 $m_{\oplus}$ for $\Sigma_0 \sim$ 1--2 
g cm$^{-2}$. Because young stars appear to have a wide range of 
initial disk masses, we expect a wide range in the masses of
terrestrial planets in exosolar systems. In future studies, we plan 
to address this issue in more detail.

In terrestrial planet formation, the transitions between
different stages of growth produce distinct waves through
the disk. During the transition from orderly growth to
runaway growth, the increase in the collision rate rapidly
propagates from the inner edge of the disk to the outer
edge. This transition is rapid and takes $\lesssim 10^5$ yr
to move from 0.4 AU to 2 AU. 

Although less rapid, the transitions from runaway to oligarchic 
growth and from oligarchic to chaotic growth also propagate 
from the inner disk to the outer disk. During the transition
to chaotic growth, dynamical interactions tend to produce more
chaotic orbits at the outer edge of the disk. This behavior
depends on the surface density gradient. In disks with
steep density gradients, $\Sigma \sim \Sigma_0 a^{-n}$ with
$n \gtrsim$ 3/2, chaotic interactions propagate slowly outward.
In disks with shallower density gradients, $n \lesssim$ 1,
dynamical interactions tend to concentrate more mass in the
outer part of the disk. This difference in behavior is set
by the growth rate, $\sim P/\Sigma \sim a^{n+3/2}$, where $P$
is the orbital period: the wave of growth propagates more 
rapidly through disks with shallower density gradients 
\citep[e.g.][]{lis87}.

These results have several interesting consequences for the 
evolution of planets in the terrestrial zone 
\citep[see also][]{kom02}. 
The transition from oligarchic to chaotic growth occurs on 
timescales, $\sim$ a few Myr, well before radiometric evidence 
suggests the formation of the Earth was fairly complete 
\citep[e.g.,][]{yin02}. 
Planets are also fully formed well before the estimated
time of the Late Heavy Bombardment, $\sim$ 100--300 Myr
after the formation of the Sun \citep[e.g.,][]{ter74, 
har80, ryd02, koe03}

Throughout the chaotic growth phase, our calculations produce
many lunar- to Mars-sized objects on highly eccentric orbits.
These objects are good candidates for the `giant impactor' 
that collided with the Earth to produce the Moon
\citep{har75b,cam76,ben86,can04a,can04b}. As we complete
calculations with fragmentation and migration, predicted 
mass and eccentricity distributions for these objects will 
yield better estimates for the probability of these events.

Together with the dynamical influence of Jupiter 
\citep[e.g.,][]{kom04}, the highly eccentric orbits of lower 
mass oligarchs have an important role in clearing the inner 
solar system \citep[e.g.,][]{gol04a,kom04} and delivering 
water to the Earth \citep{lun03}.  Traditionally, data from 
our solar system provide the only tests of clearing mechanisms 
\citep[e.g.,][and references therein]{gro01,nes02a,nes02b}.
In the next decade, comparisons between predicted infrared 
excesses from the debris disks leftover from terrestrial 
planet formation and observations from {\it Spitzer} and 
{\it TPF-Darwin} will yield new constraints on clearing 
timescales and the evolution of volatile species in the 
terrestrial zone \citep[e.g.,][]{bei05}. These comparisons 
will enable better numerical calculations and an improved 
understanding of the final stages of terrestrial planet 
formation.

\acknowledgements

We thank M. Geller and an anonymous referee for important comments 
that improved the content of this paper.  We acknowledge a generous 
allotment, $\sim$ 5 cpu years, of computer time on the Silicon 
Graphics Origin-2000 computers `alhena', `castor', and `pollux' 
and the Dell Linux cluster `cosmos' at the Jet Propulsion Laboratory 
through funding from the JPL Institutional Computing and Information
Services and the NASA Directorates of Aeronautics Research, Science,
Exploration Systems, and Space Operations.  We thank V. McGlasson, 
M. Phelps, and other staff members of the CfA computation facility for 
installation and support of the SUN cluster `hydra', where we used 
$\sim$ 6 cpu years to perform many of our calculations.  The {\it NASA} 
{\it Astrophysics Theory Program} supported part of this 
project through grant NAG5-13278.

\clearpage

\begin{deluxetable}{cccc}
\tablecolumns{4}
\tablewidth{0pc}
\tabletypesize{\normalsize}
\tablenum{1}
\tablecaption{Transition from Oligarchy to Chaos at 0.84--1.16 AU}
\tablehead{
\colhead{$\Sigma_0$ (g cm$^{-2}$)} & 
\colhead{log $t_m$ (yr)} &
\colhead{$r_{\Sigma}$} &
\colhead{$n_{n,max}$} }
\startdata
~1 & 7.0 $\pm$ 0.05 & 0.58 $\pm$ 0.01 & 0.85 $\pm$ 0.05 \\
~2 & 6.6 $\pm$ 0.05 & 0.53 $\pm$ 0.01 & 0.91 $\pm$ 0.05 \\
~4 & 6.3 $\pm$ 0.07 & 0.48 $\pm$ 0.02 & 0.99 $\pm$ 0.10 \\
~8 & 6.0 $\pm$ 0.08 & 0.41 $\pm$ 0.02 & 1.11 $\pm$ 0.07 \\
12 & 5.7 $\pm$ 0.10 & 0.32 $\pm$ 0.03 & 1.24 $\pm$ 0.12 \\
16 & 5.3 $\pm$ 0.10 & 0.24 $\pm$ 0.03 & 1.33 $\pm$ 0.11 \\
\enddata
\end{deluxetable}
\clearpage


\begin{figure}
\plotone{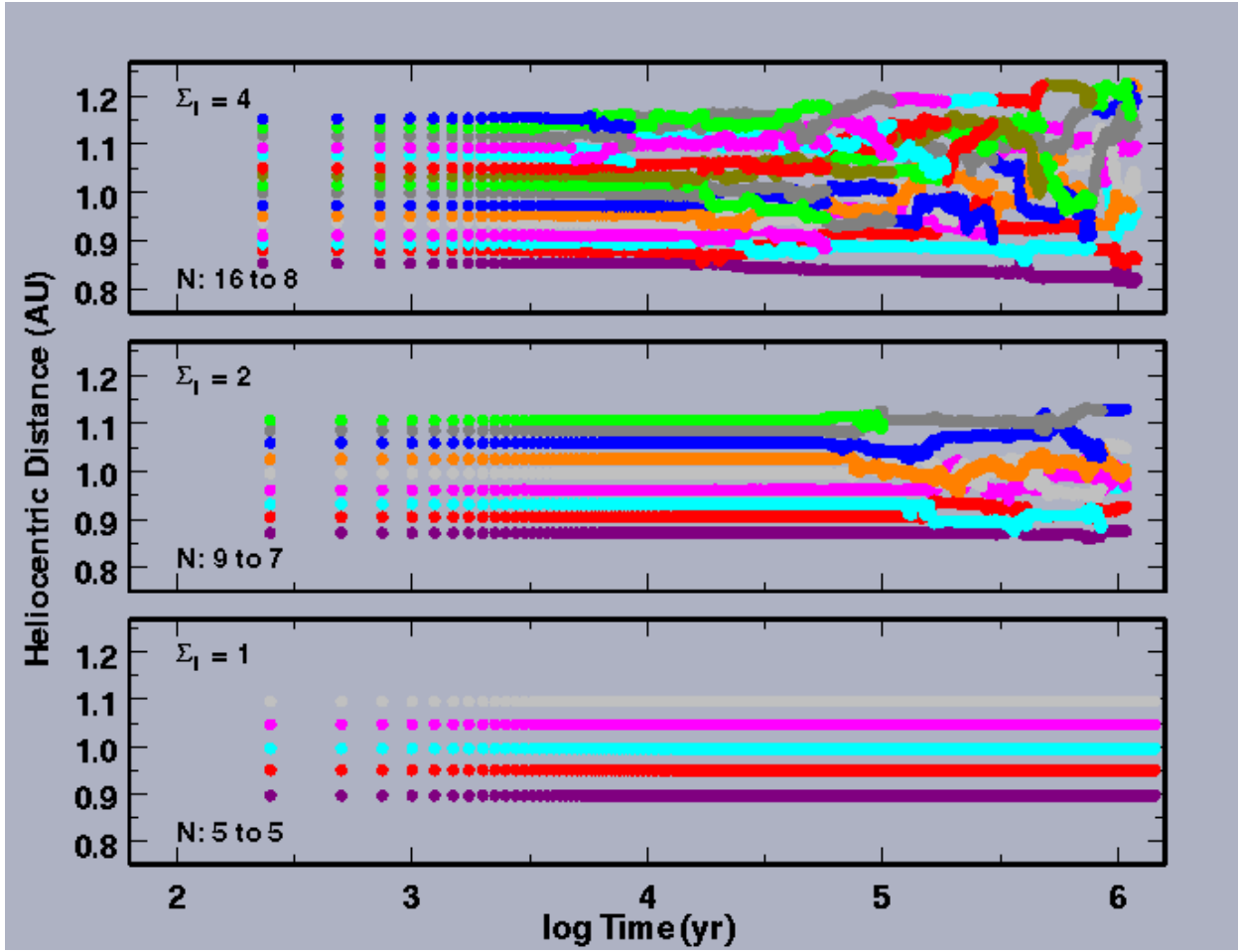} 
\caption
{Evolution of eccentricity for oligarchs in a swarm of
lower mass planetesimals. Each colored track shows the
eccentricity of a single oligarch with time. A change in 
the color of a track (for example, in the middle panel at
1 Myr when the dark gray track turns blue) indicates a merger 
of one pair of oligarchs. The legend of each panel lists the 
surface density $\Sigma_l$ in g cm$^{-2}$ at 1 AU of oligarchs 
with individual masses $m_s = 10^{26}$ g.  In each panel, the 
surface density $\Sigma_s$ of planetesimals with masses $m_s =$
$10^{24}$ g equals the surface density of oligarchs. When 
$\Sigma_l \lesssim$ 1 g cm$^{-2}$, the orbits of oligarchs
do not overlap on timescales of 1 Myr. For $\Sigma_l \gtrsim$
2 g cm$^{-2}$, gravitational stirring leads to overlapping orbits
and mergers of oligarchs. More massive systems develop
overlapping orbits on shorter timescales, which leads to more
mergers on timescales of 1 Myr or less.}
\end{figure}

\begin{figure}
\plotone{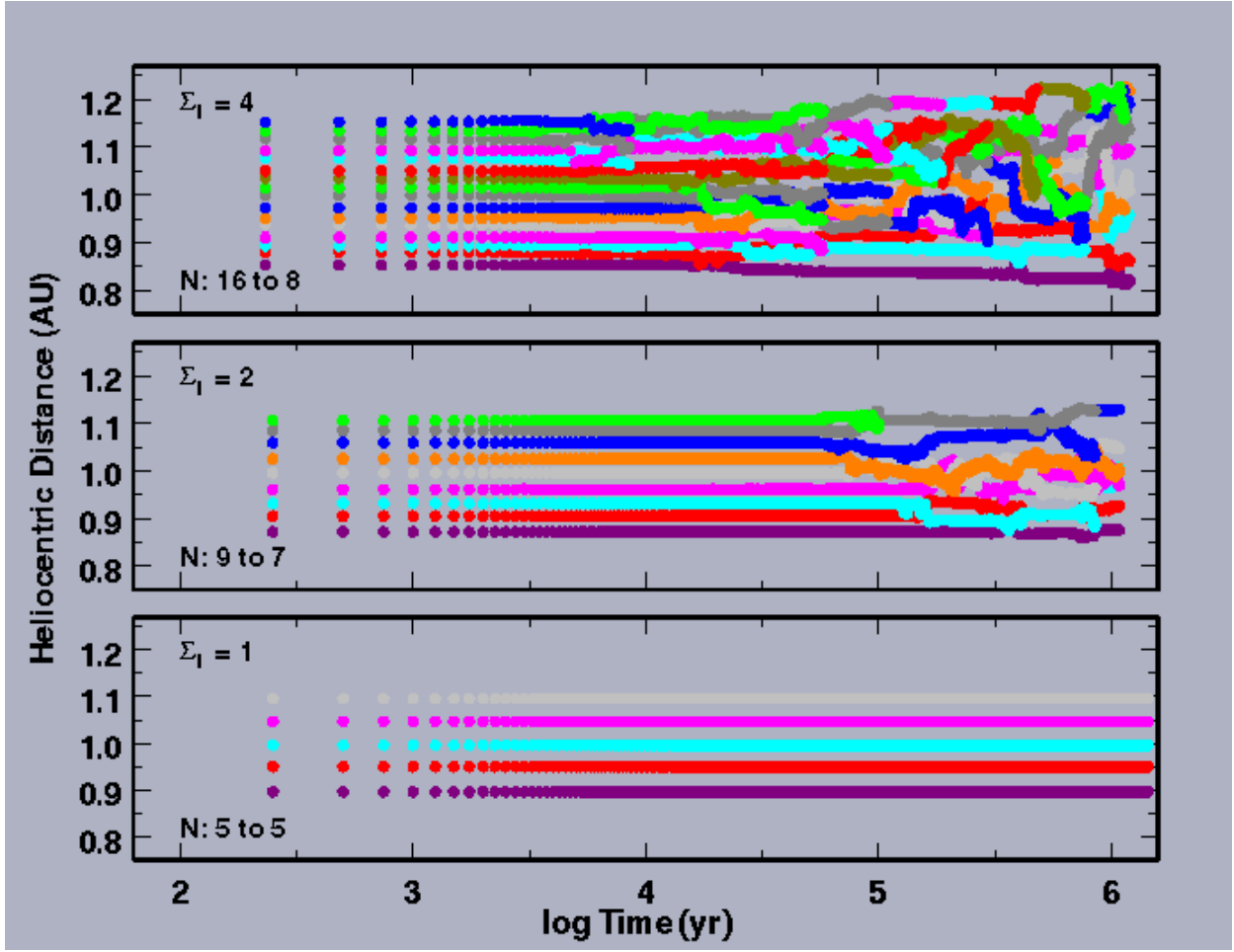} 
\caption  
{As in Figure 1 for the semimajor axes of the oligarchs.
The legend in the lower left corner of each panel indicates
the number of oligarchs at the beginning and end of each 
calculation.}
\end{figure}

\begin{figure}
\plotone{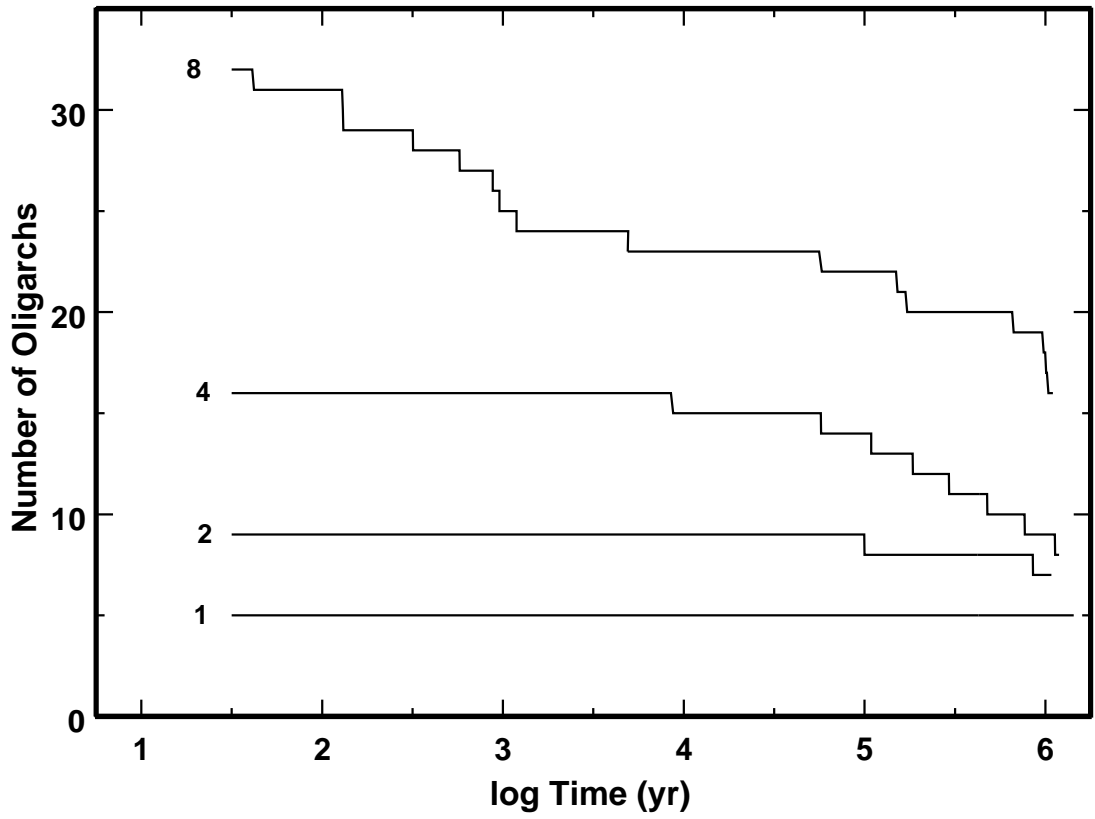} 
\caption
{Evolution of the number of oligarchs for models with 
$\Sigma_l$ = 1--8 g cm$^{-2}$.  
More massive systems produce more mergers on shorter timescales.}
\end{figure}

\begin{figure}
\plotone{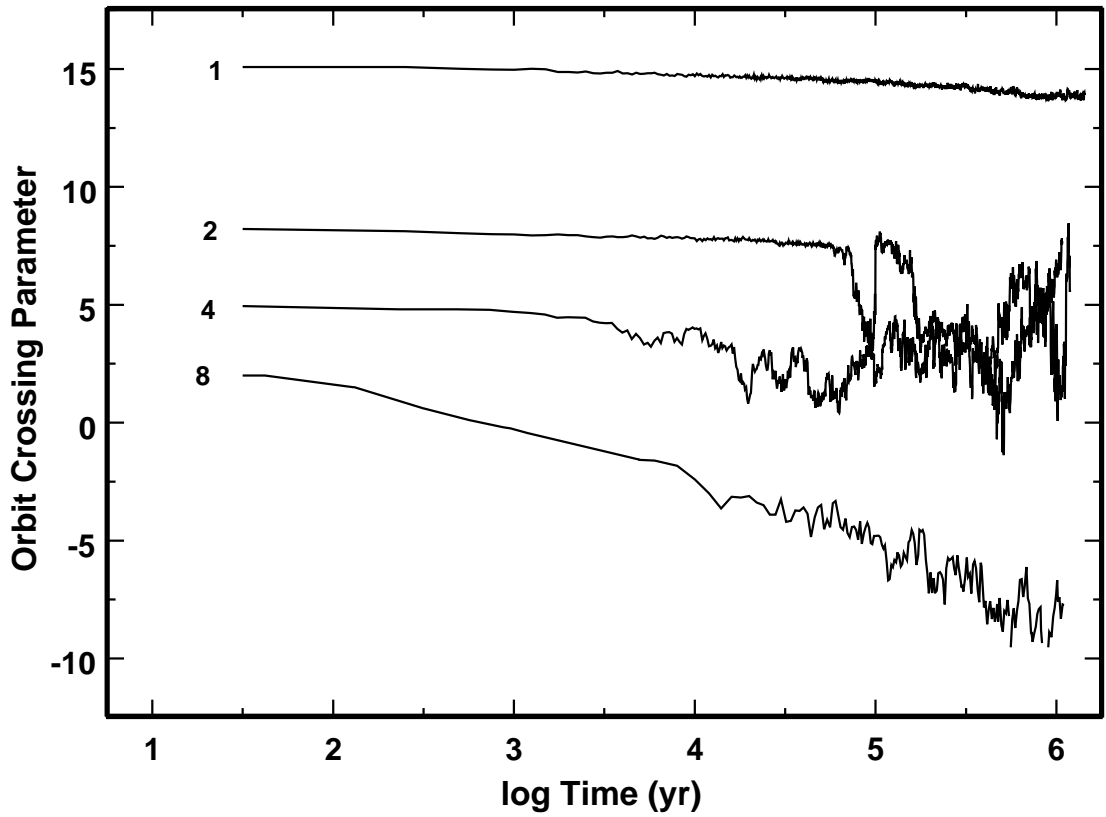} 
\caption
{Evolution of the orbit crossing parameter $p_o$ for models 
with $\Sigma_l$ = 1--8 g cm$^{-2}$. For $p_o \gtrsim 2$, 
orbits of oligarchs do not overlap and mergers do not occur.
As $p_o$ approaches 0, overlapping orbits of larger objects 
lead to mergers.}
\end{figure}

\begin{figure}
\plotone{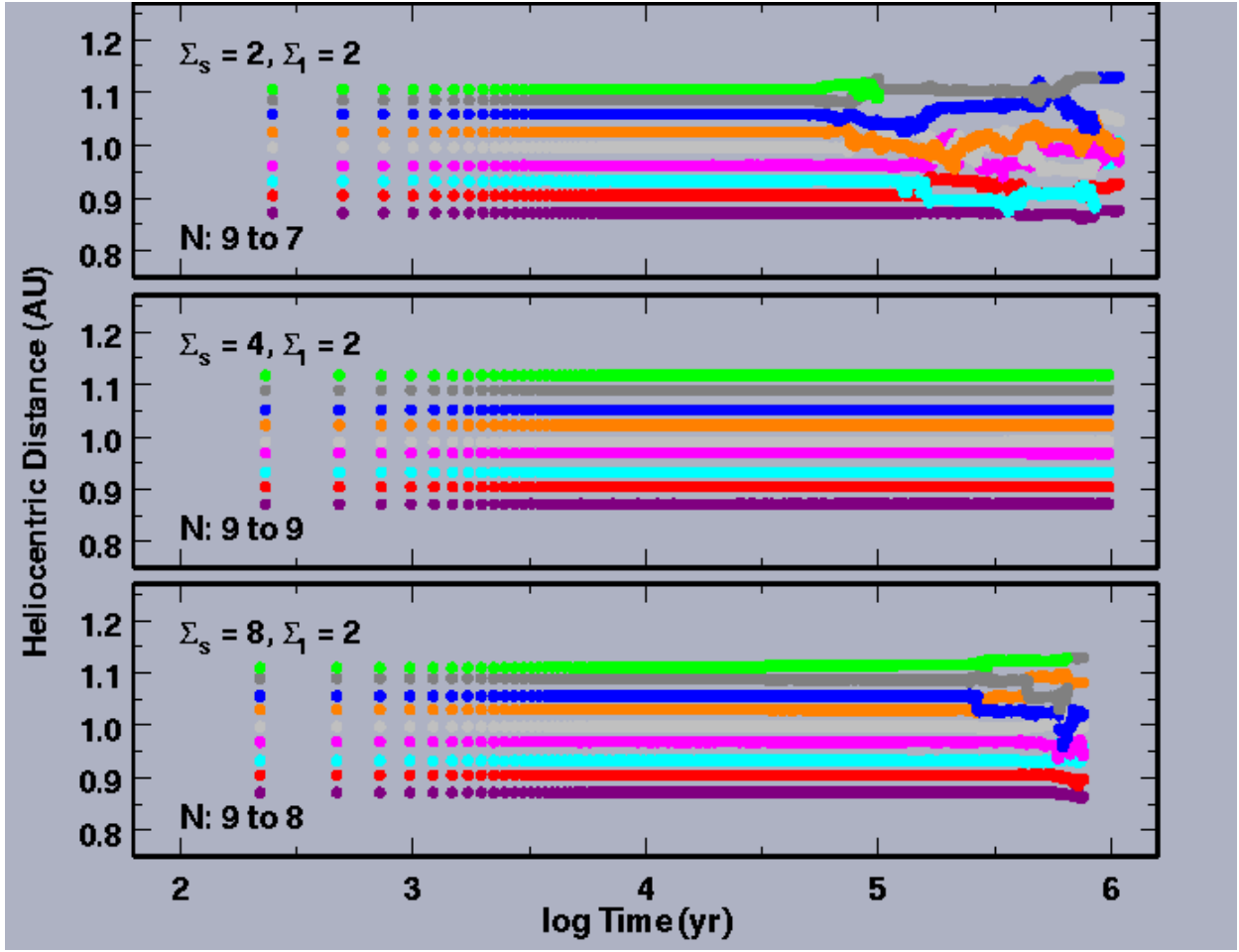} 
\caption
{Evolution of semimajor axes for oligarchs in a swarm of
lower mass planetesimals. As in Figure 1, each colored track
show the position of one oligarch; changes in color along a 
track indicate a merger. The legend of each panel lists
the number of oligarchs and the surface densities of 
planetesimals with $m_s = 10^{24}$ g ($\Sigma_s$) and
oligarchs with $m_l = 10^{26}$ g ($\Sigma_l$) at 1 AU. 
As more mass is placed in smaller objects, dynamical friction
first reduces orbit overlap between oligarchs. However,
when the planetesimals contain more mass, viscous stirring 
between planetesimals aids orbit overlap and leads to 
mergers of the oligarchs.}
\end{figure}

\begin{figure}
\plotone{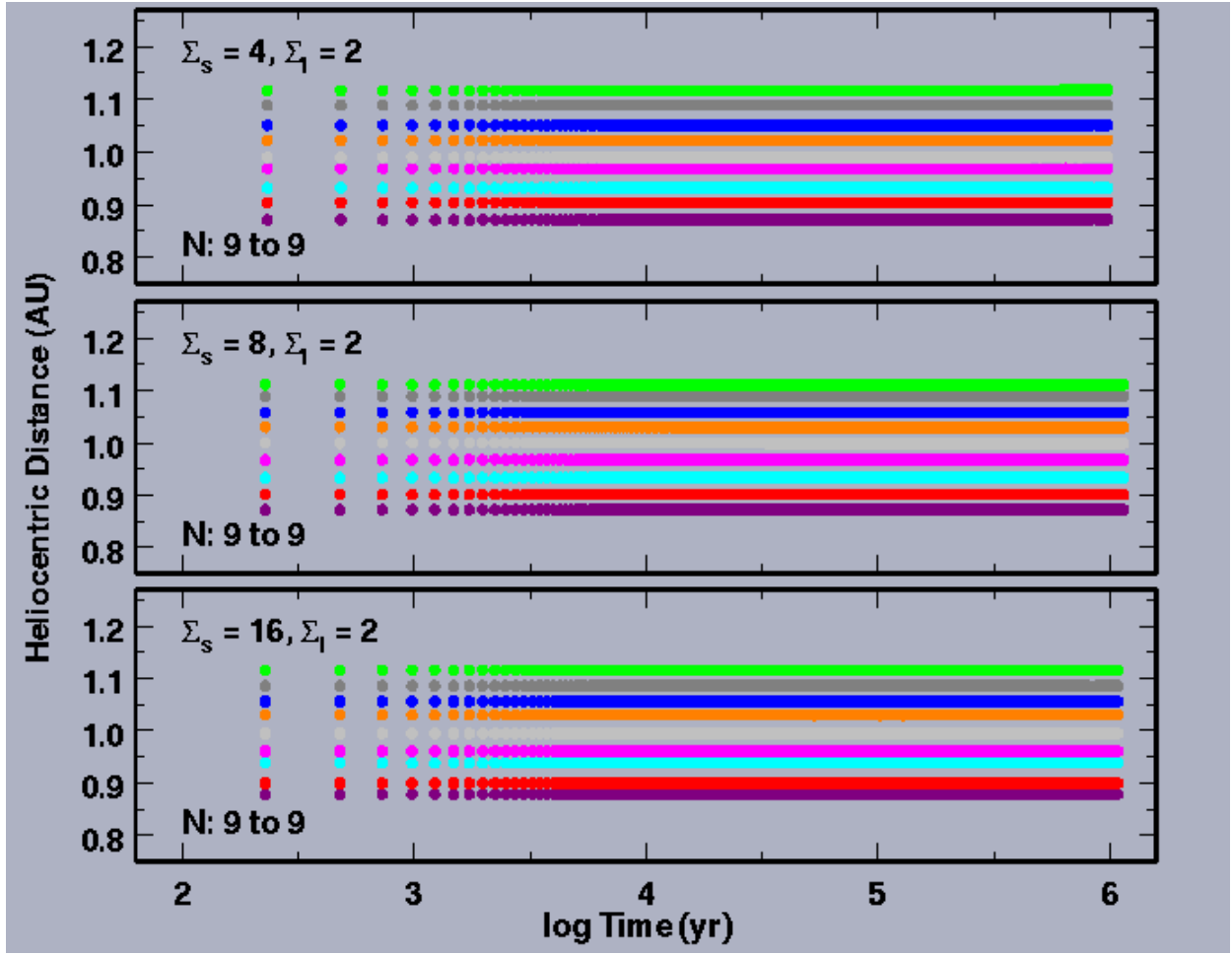} 
\caption
{As in Figure 4 for $m_s = 10^{16}$ g. Viscous stirring 
between planetesimals is minimal and does not lead to
overlapping orbits of oligarchs in 1 Myr.}
\end{figure}

\begin{figure}
\plotone{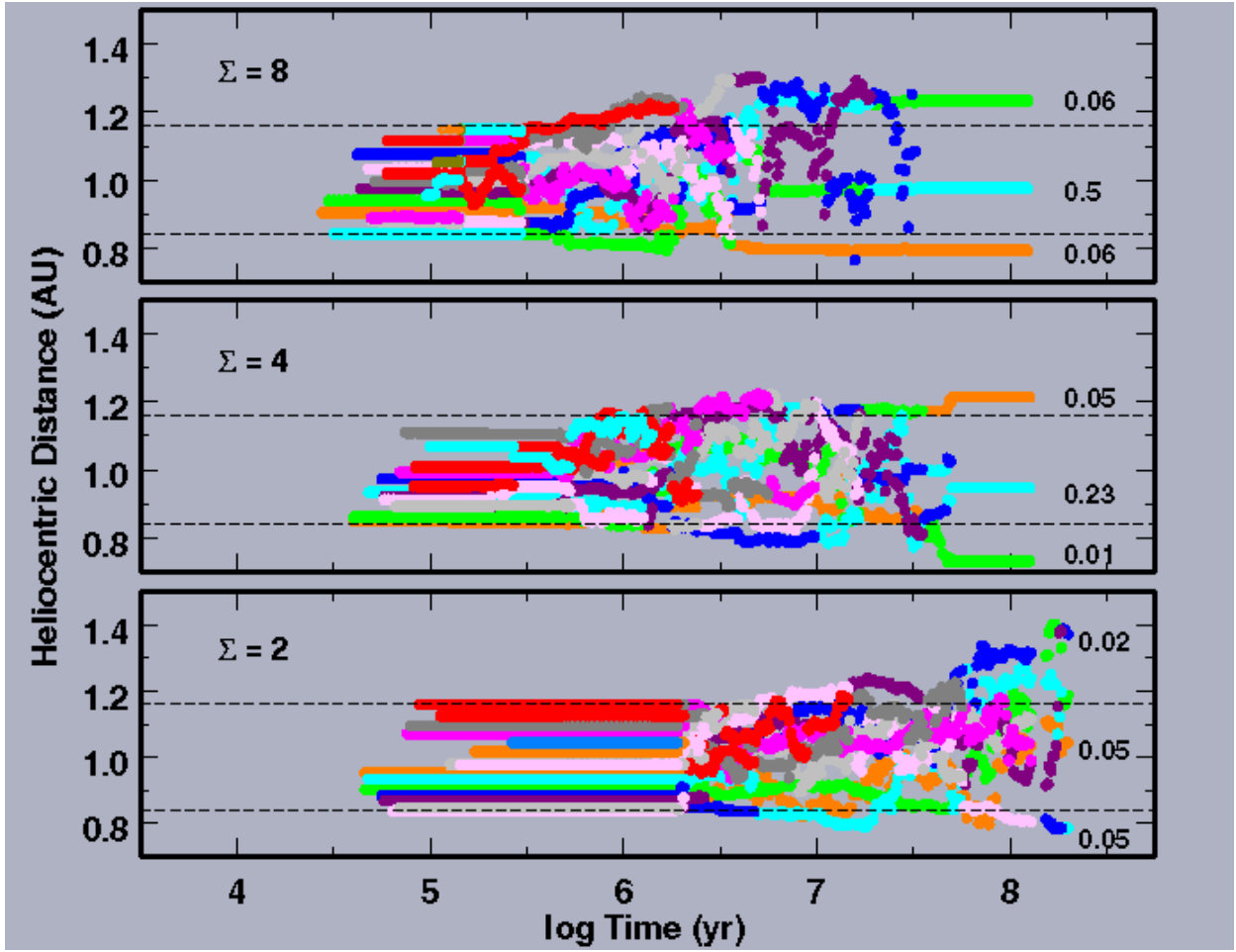} 
\caption
{Evolution of semimajor axes for oligarchs with masses larger than 
$m_{pro} = 10^{25} ~ (\Sigma_0 / 4~{\rm g~cm^{-2}})$ g in 
a full planet formation calculation at 0.86--1.14 AU. The 
legend of each panel lists the initial surface density in 
1--10 km planetesimals at 1 AU (upper left corner, in units 
of g cm$^{-2}$), and the mass of the largest planets (in 
$m_{\oplus}$; at the right end of each track). 
The dashed lines indicate the extent of the planetesimal grid 
in the coagulation code. Runaway growth produces protoplanets 
in $10^4$ yr to $10^5$ yr. Continued growth of protoplanets 
during the oligarchic phase leads to orbit overlap on timescales 
of roughly 1 Myr. More massive systems reach this limit faster 
than less massive systems. Overlapping orbits leads to merger 
and more rapid growth of protoplanets.}
\end{figure}

\begin{figure}
\plotone{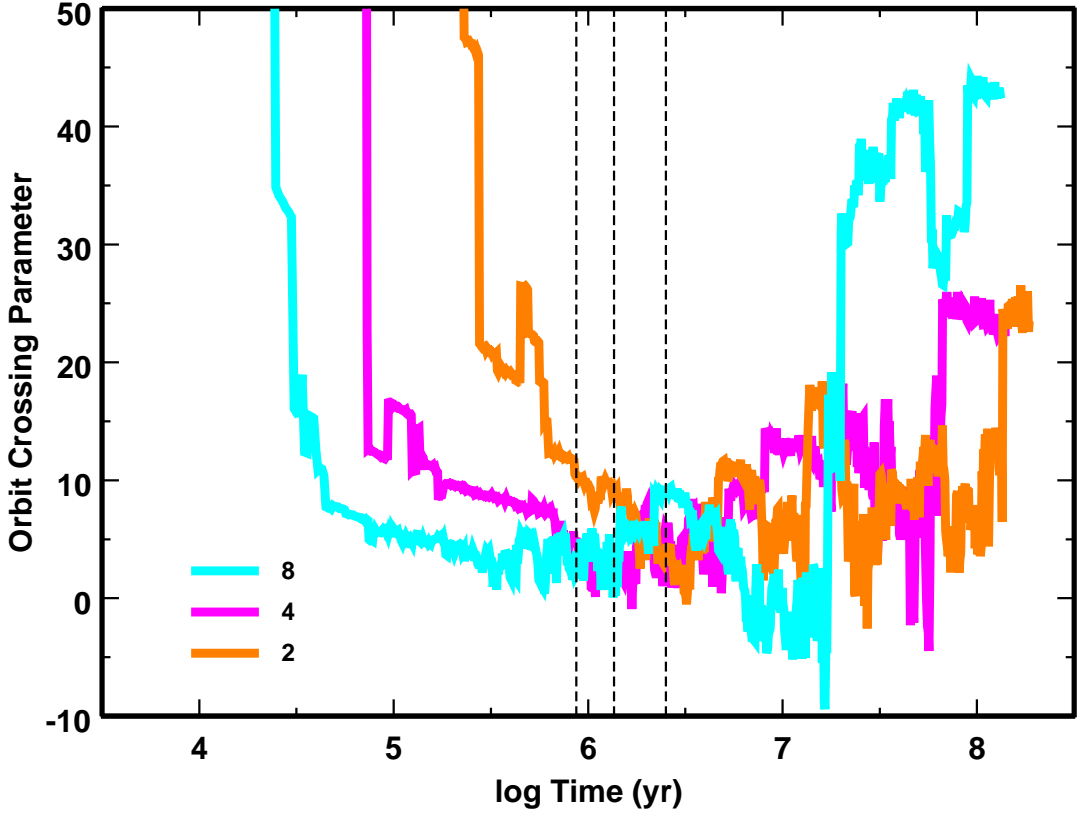} 
\caption
{Evolution of the orbit crossing parameter $p_o$ for the models 
of Figure 7. The legend lists the surface density in units of 
g cm$^{-2}$ for each model. The dashed lines indicate when the 
surface density of oligarchs is half the total surface density. 
The leftmost dashed line corresponds to the largest initial 
surface density; the rightmost line corresponds to the smallest
initial surface density.  As oligarchs form, $p_o$ declines 
and reaches a plateau where $p_o \approx$ 1--5. Continued 
growth of the oligarchs leads to a minimum in $p_o$, roughly when 
$\Sigma_l \approx \Sigma_s$.  minima Subsequent mergers reduce $p_o$ 
until the system is in rough equilibrium, when $p_o$ increases
dramatically.}
\end{figure}

\begin{figure}
\plotone{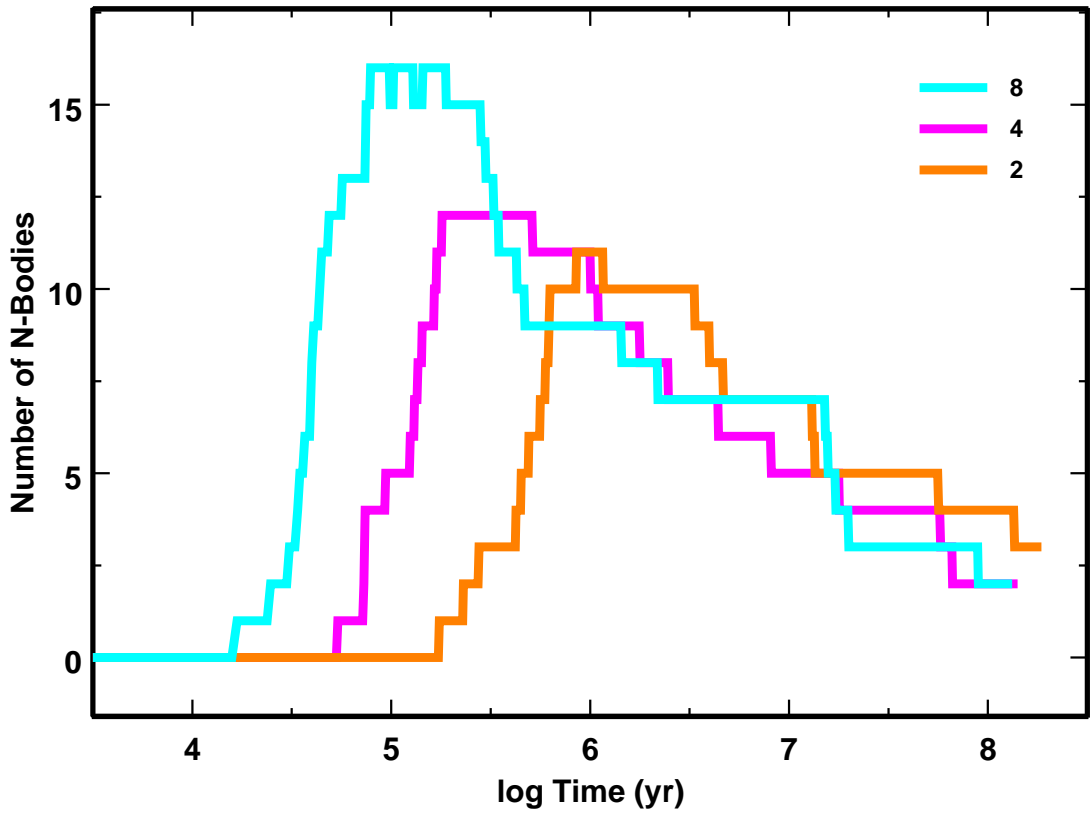} 
\caption
{Evolution of the number of oligarchs $N_o$ for the models of Figure 7.}
\end{figure}

\begin{figure}
\plotone{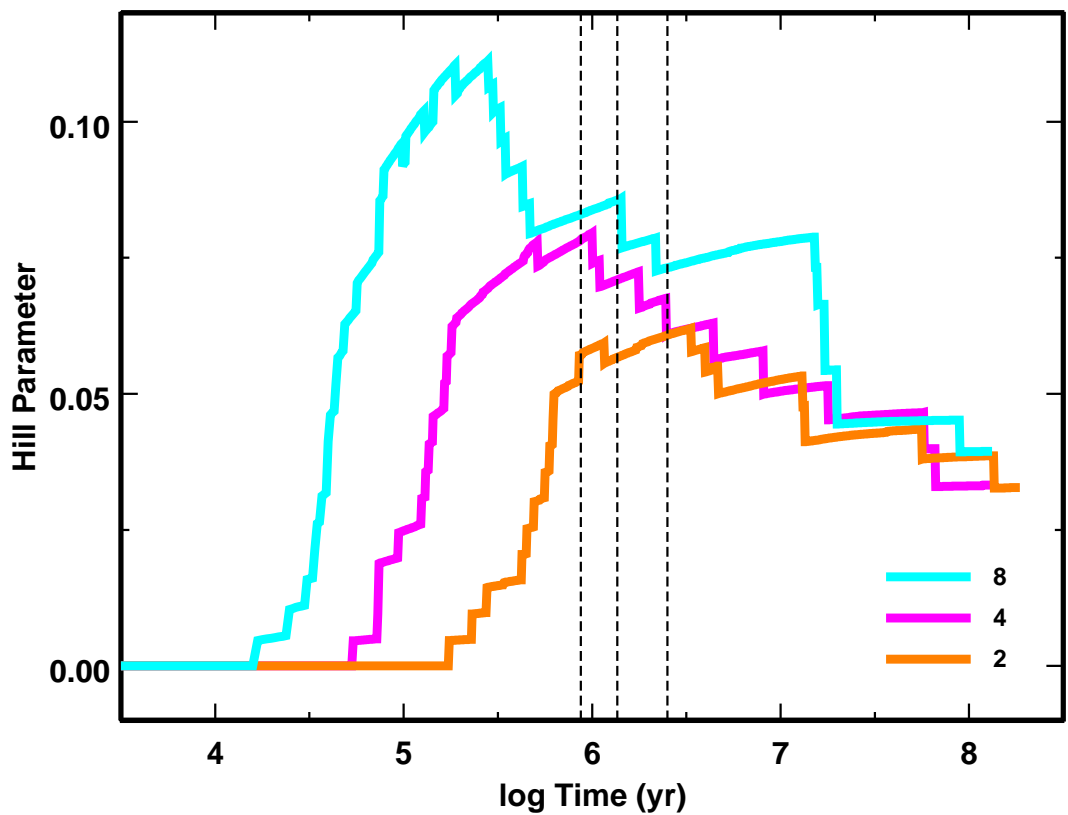} 
\caption
{Evolution of the Hill parameter $p_H$ for the models of Figure 7.
The dashed lines indicate when the surface density of oligarchs 
is half the total surface density. The leftmost dashed line 
corresponds to the largest initial surface density; the rightmost 
line corresponds to the smallest initial surface density.  
When the number of oligarchs is maximum, the Hill parameter 
peaks. Mergers reduce $p_H$; growth increases $p_H$. Chaotic
growth occurs when $p_H \gtrsim$ 0.07--0.1.}
\end{figure}

\begin{figure}
\plotone{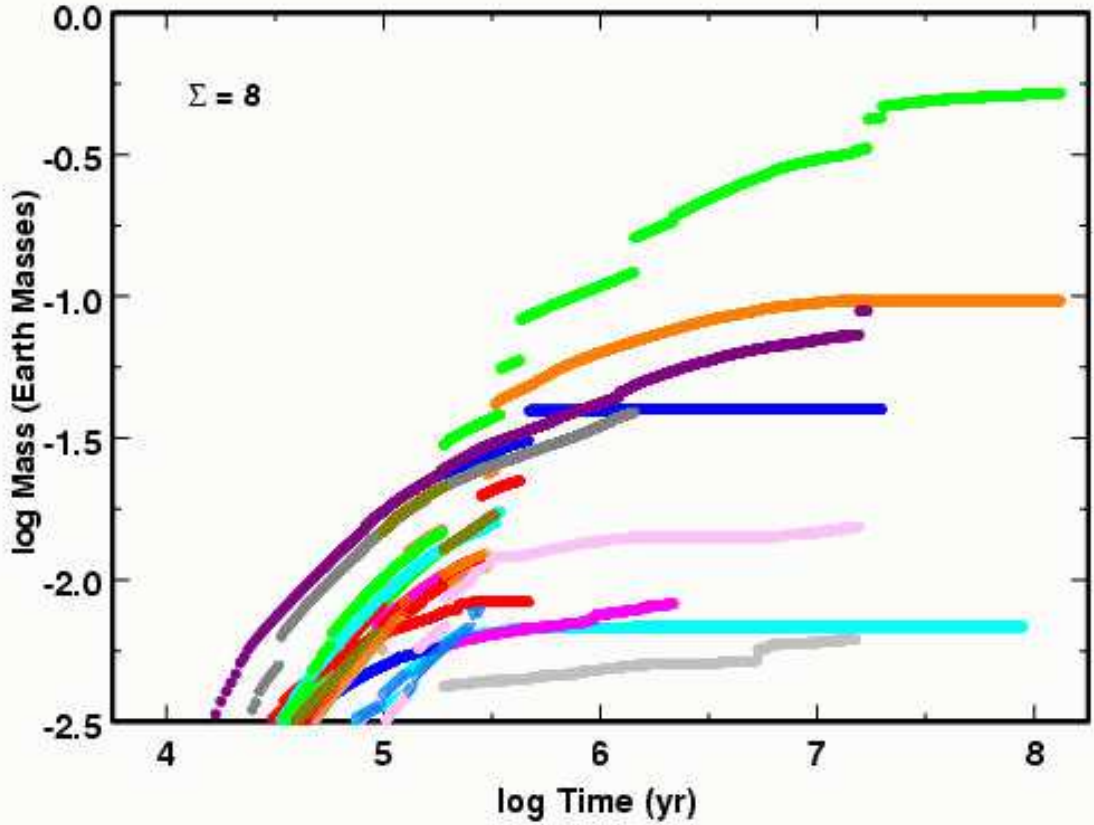}
\caption
{Variation of $t_{\Sigma}$ and $t_m$ as a function of $\Sigma_0$ 
for different values of $m_{pro}$. The legend indicates symbols 
for the logarithm of the promotion mass.  For each promotion mass 
and surface density, the points indicate typical values and extreme 
values. Thus, the two panels provide a visual impression of the 
dispersion of the results among a large set of calculations.
} 
\end{figure}

\begin{figure}
\plotone{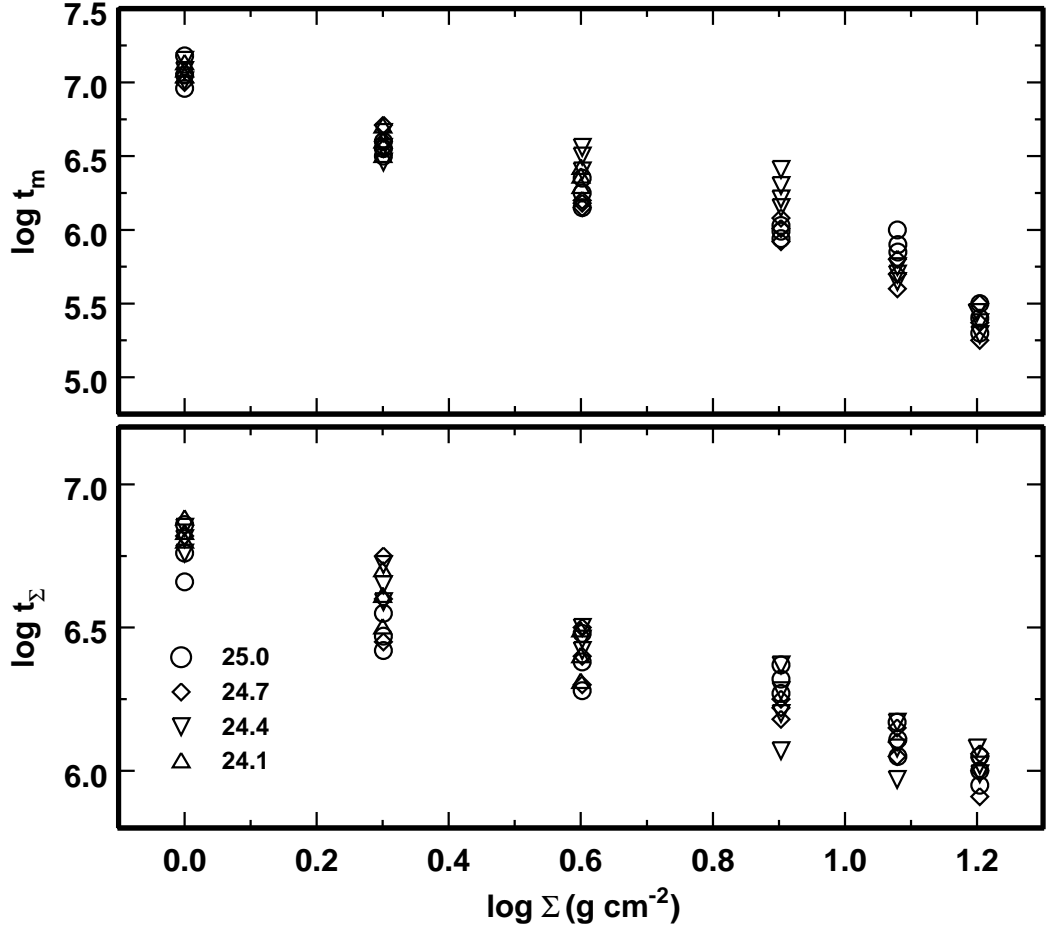} 
\caption
{Evolution of masses for oligarchs with masses larger than 
$2 \times 10^{25}$ g in a full planet formation calculation 
at 0.86--1.14 AU. The legend lists the initial surface density 
in 1--10 km planetesimals at 1 AU (upper left corner, in units 
of g cm$^{-2}$).  Each colored track shows the mass evolution
for one oligarch. Discontinuities or terminations in the tracks 
indicate mergers of large objects.} 
\end{figure}

\begin{figure}
\plotone{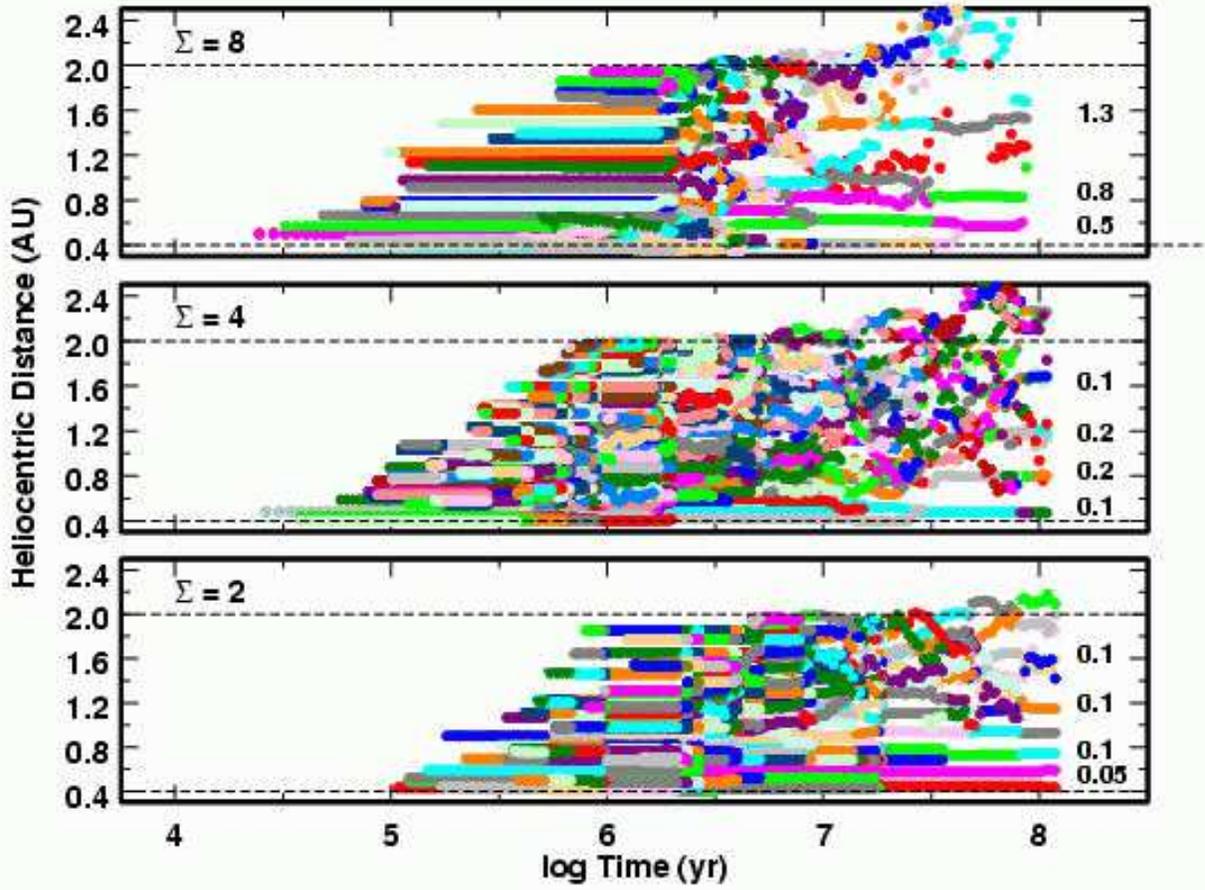} 
\caption
{As in Figure 7 for a calculation at 0.4--2 AU
with $m_{pro}$ = $2 \times 10^{25} (\Sigma/{\rm 4~g~cm^{-2}})$ g.}
\end{figure}

\begin{figure}
\plotone{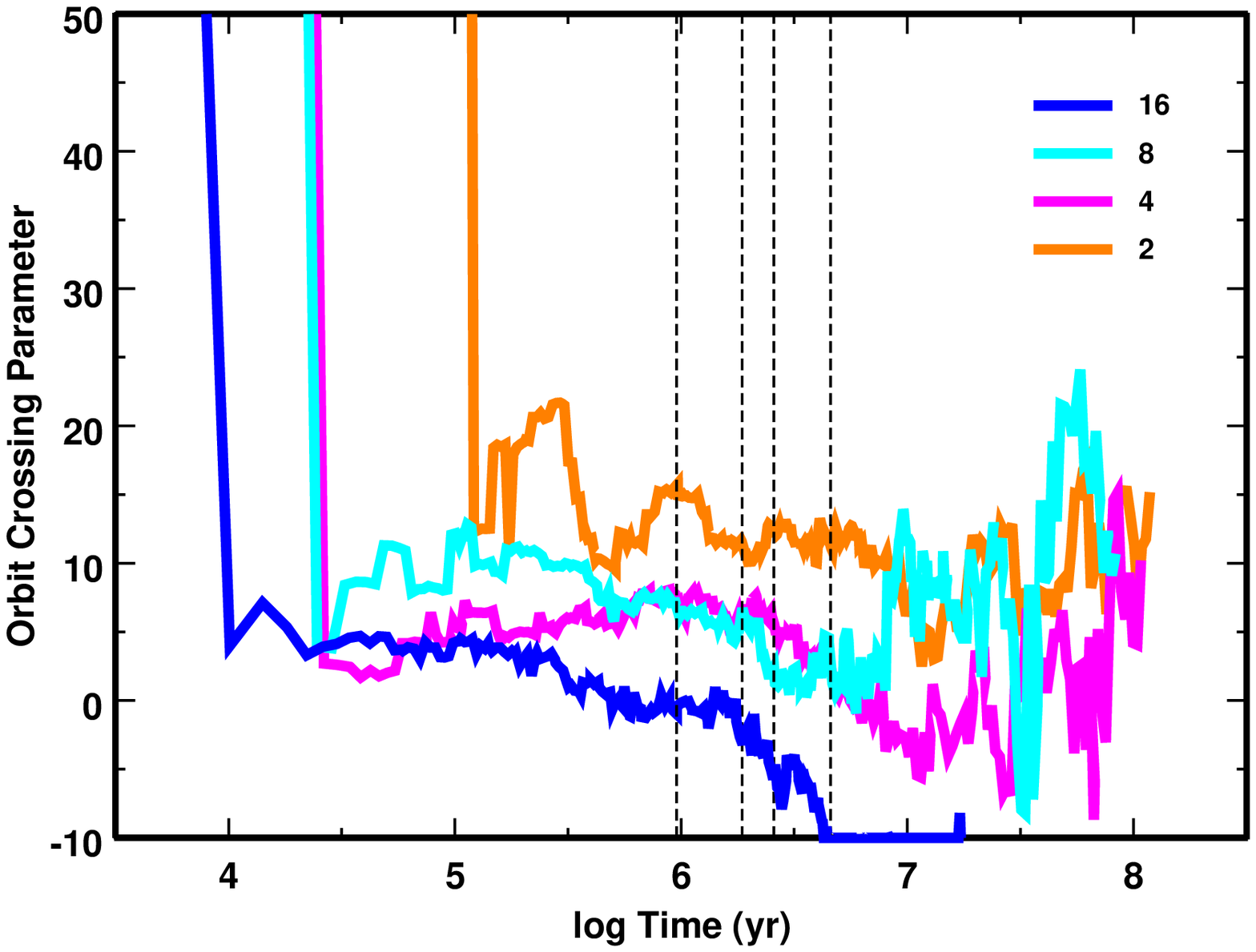} 
\caption
{As in Figure 8 for the models of Figure 13.} 
\end{figure}

\begin{figure}
\plotone{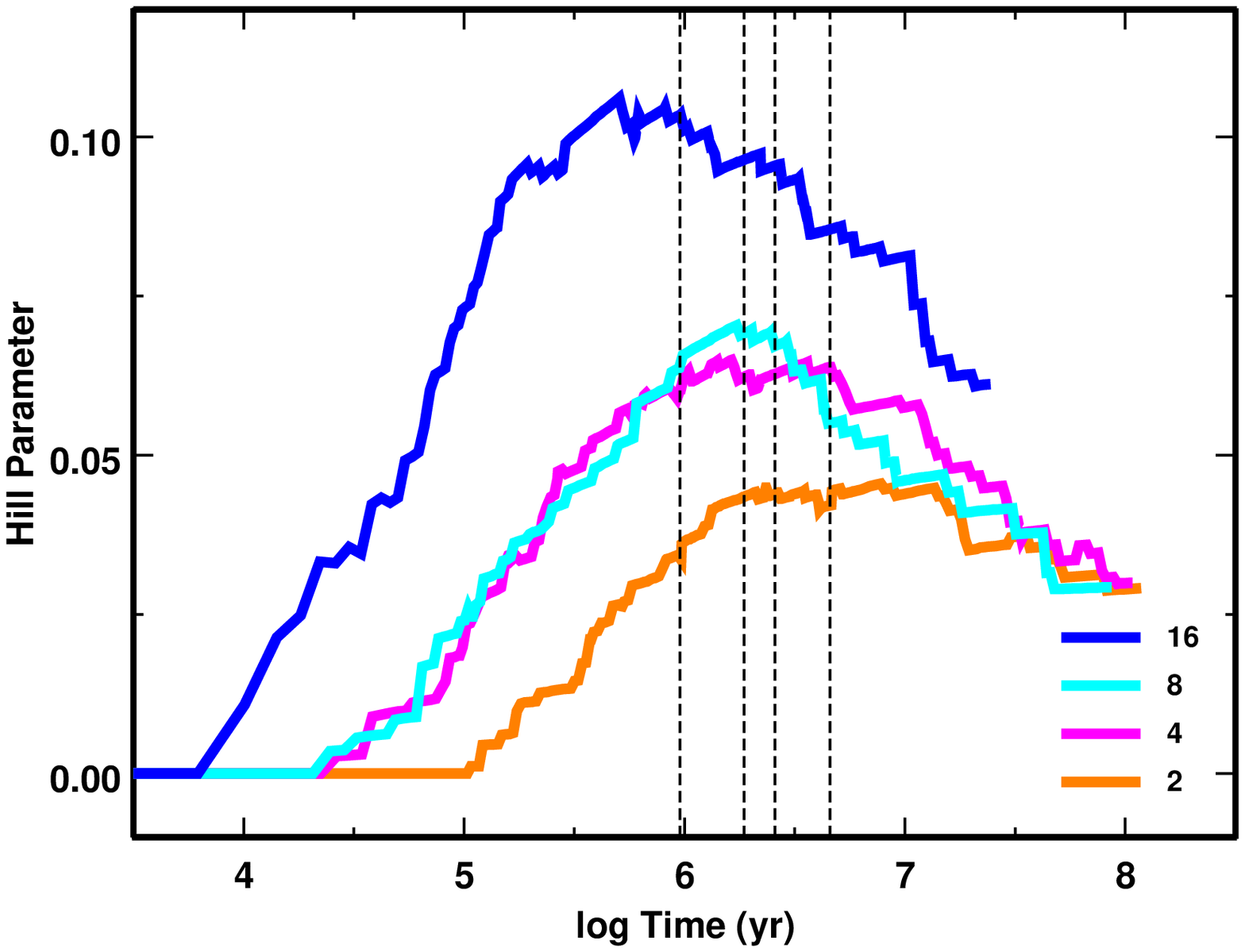} 
\caption
{As in Figure 10 for the models of Figure 13.}
\end{figure}

\begin{figure}
\plotone{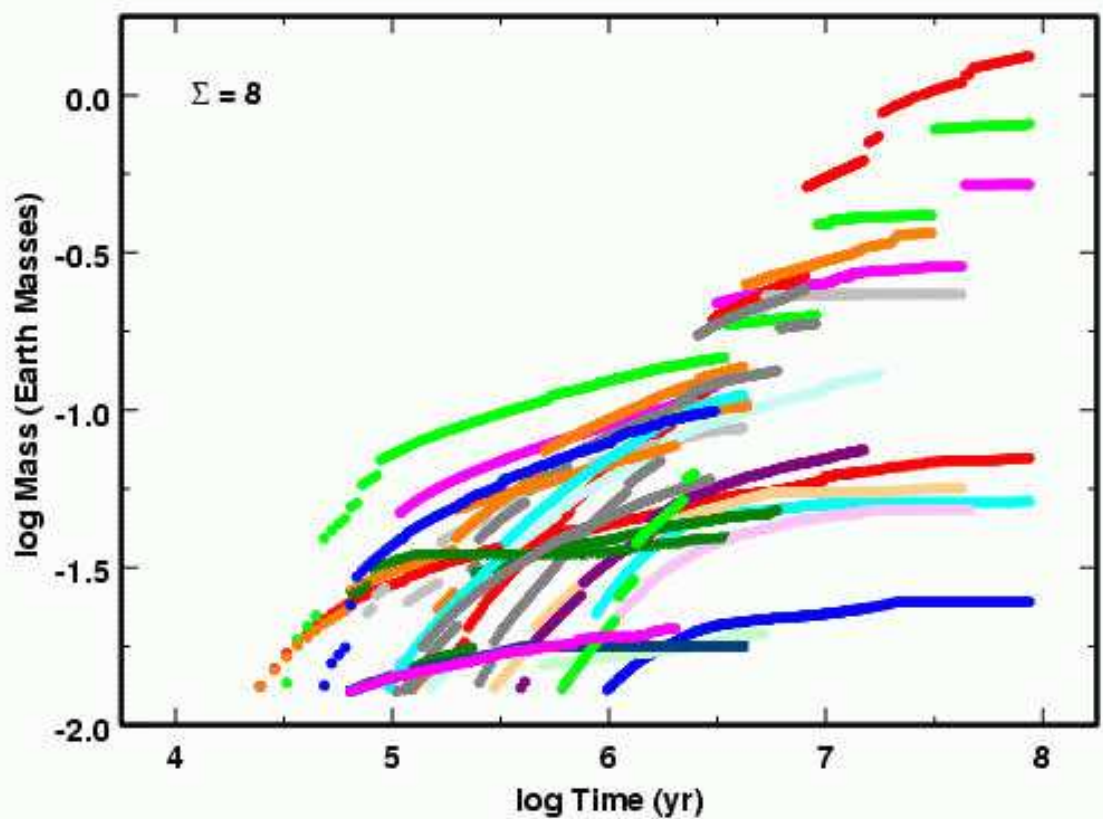} 
\caption
{As in Figure 12 for a model from Figure 13.}
\end{figure}

\clearpage

\end{document}